\documentclass[a4paper,11pt]{article}
\usepackage{amsmath,amssymb,bm}
\usepackage{titlesec}
\titleformat{\paragraph}[runin]{\normalfont\itshape}{\theparagraph.}{.3em}{}[.]\titlespacing{\paragraph}{0pt}{1ex plus .1ex minus .2ex}{.5em}
\usepackage[letterpaper, hmargin=1in, top=1in, bottom=1.2in, footskip=0.6in]{geometry}
\usepackage[utf8]{inputenc}
\usepackage{lmodern}
\usepackage{mathtools}
\usepackage{dsfont}
\pdfoutput=1
\usepackage[T1]{fontenc}
\usepackage[english]{babel}
\usepackage{graphicx} 
\usepackage{booktabs} 
\usepackage{color}
\definecolor{aquamarine}{rgb}{0.5, 1.0, 0.83}
\definecolor{ao(english)}{rgb}{0.0, 0.5, 0.0}
\definecolor{armygreen}{rgb}{0.29, 0.33, 0.13}
\definecolor{awesome}{rgb}{1.0, 0.13, 0.32}
\definecolor{ballblue}{rgb}{0.13, 0.67, 0.8}
\definecolor{bittersweet}{rgb}{1.0, 0.44, 0.37}
\definecolor{blue}{rgb}{0.0, 0.0, 1.0}
\definecolor{brinkpink}{rgb}{0.98, 0.38, 0.5}
\definecolor{ballblue}{rgb}{0.13, 0.67, 0.8}
\definecolor{brightturquoise}{rgb}{0.03, 0.91, 0.87}
\definecolor{blue-green}{rgb}{0.0, 0.87, 0.87}
\definecolor{caribbeangreen}{rgb}{0.0, 0.8, 0.6}
\definecolor{cyan}{rgb}{0.0, 1.0, 1.0}
\definecolor{amber(sae/ece)}{rgb}{1.0, 0.49, 0.0}
\graphicspath{ {images/} }

\author{J\"urg Fr\"ohlich\footnote{Email: juerg@phys.ethz.ch}\\
Institute for Theoretical Physics\\
ETH Zurich\\
8093 Zurich, Switzerland}

\title{Phase Transitions, Spontaneous Symmetry Breaking, and Goldstone's Theorem\footnote{published in: 
Encyclopedia of Condensed Matter Physics, Second Edition, T. Chakraborty (ed.), Elsevier 2024, https://doi.org/10.1016/B978-0-323-90800-9.00275-4}}

\begin{document}

\maketitle

\begin{abstract}
Some important rigorous results on phase transitions accompanied by the spontaneous breaking of 
symmetries in statistical mechanics and relativistic quantum field theory are reviewed.
Basic ideas, mainly inspired by quantum field theory, underlying the proofs of some of these results 
are sketched. The Goldstone theorem is proven, and the Mermin-Wagner-Hohenberg theorem concerning 
the absence of continuous symmetry breaking in one and two dimensions is recalled. Comments 
concerning rigorous results on the Kosterlitz-Thouless transition in the two-dimensional classical XY model 
are made.
\end{abstract}


\noindent
Key Points:
\begin{itemize}
\item{A heuristic introduction to the subject of spontaneous symmetry breaking, Goldstone's theorem 
and the Anderson-Higgs mechanism in the context of quantum field theory is given.
The Wick rotation from real to imaginary time and the equivalence of real-time relativistic 
quantum field theory (RQFT) and imaginary-time Euclidean field theory (EFT) are recalled.}
\item{The $\lambda |\phi|^{4}$-Euclidean field theory of a real or complex scalar field $\phi$ 
in three dimensions is introduced. Its relevance for understanding various systems of
statistical mechanics and condensed matter physics, e.g, Bose gases, is commented upon. 
It is shown that the broken-symmetry phase of the theory exhibits massless modes, in accordance 
with the Goldstone theorem.}
\item{A general proof of the Goldstone theorem in the formalism of EFT is presented. The example
of $\lambda |\phi|^{4}$-theory of a complex scalar field in three dimensions is considered in more
detail. Bounds on critical exponents for the susceptibility of the theory and its correlation length, 
as phase transition points in zero magnetic field are approached, are reviewed.}
\item{A summary of results on phase transitions accompanied by spontaneous symmetry breaking
for classical and quantum (lattice) spin systems is presented. The role of ``Reflection Positivity'' and 
``Infrared Bounds'' in the analysis of phase transitions is emphasized. For the classical XY model, 
the usefulness of a representation as a vortex gas for the analysis of the massless low-temperature phase 
and for the analysis of the Kosterlitz-Thouless transition is indicated.}
\end{itemize}

\section{Introduction to spontaneous symmetry breaking and the Higgs mechanism}\label{Intro}

This contribution is \textbf{not} an exhaustive review of the subject of spontaneous breaking of (continuous) 
symmetries or its absence, the Goldstone theorem, the Mermin-Wagner-Hohenberg theorem, 
the Anderson-Higgs mechanism, etc. My modest goal is to provoke the readers' interest in this subject,
to draw attention to the fruitful intertwining between statistical mechanics and particle physics,
in particular quantum field theory, and to draw some attention to the more mathematical literature on these 
matters. I will not state the relevant results in all detail, let alone describe proofs. For further information
the readers are invited to consult the papers quoted in the text.

Phase transitions in systems of condensed matter have been studied ever since the $19^{th} $ century,
using the formalism of thermodynamics, by \textit{Clausius, Clapeyron, Maxwell, Gibbs} and others. Their
findings are well known and are taught in every introductory course on the theory of heat. Towards the 
end of the $19^{th}$ and at the beginning of the $20^{th}$ century, statistical mechanics was
created by Maxwell, \textit{Boltzmann,} Gibbs and \textit{Einstein.} The first example of a phase transition accompanied
by the breaking of a continuous symmetry, treated by using the methods of equilibrium statistical mechanics,
was \textit{Bose-Einstein condensation} exhibited by ideal Bose gases when the temperature is lowered
at some fixed positive density. Further examples relevant for the description of liquid-gas transitions 
and anisotropic magnetism, mostly with discrete symmetries, were studied subsequently. A famous 
model in this context is the Ising model. The phenomena of superconductivity and ferromagnetism
inspired many further efforts to understand phase transitions, spontaneous symmetry breaking and 
critical phenomena near continuous transitions. It is not the purpose of this contribution 
to provide a detailed overview on this field. General surveys of equilibrium statistical mechanics,
including the theory of phase transitions, can be found in the monographs \cite{Ruelle, B-Rob, BS}, which are
written in the style of mathematical physics. Rich sources of results on continuous phase transitions, 
symmetry breaking and critical phenomena are \cite{DGJ} and \cite{D-Itz}. 
Phase transitions accompanied by spontaneous symmetry breaking and the Anderson-Higgs 
mechanism later came to play a fundamental role in particle physics; see \cite{Weinberg} and 
references given there. As already mentioned, a very attractive feature of this subject is that it is a 
great example of productive interactions between statistical and condensed matter physics, 
on one hand, and particle physics and quantum field theory, on the other hand.

I begin my tale by briefly recalling the heuristics of spontaneous symmetry breaking and the Anderson-Higgs 
mechanism in relativistic quantum field theory (RQFT). The simplest theory is one of a complex scalar 
field $\phi$, which I write in polar coordinates
\begin{equation}\label{Einstein}
\phi(x)= e^{i\theta(x)}\,\rho(x)\,,\qquad x\in \mathbb{M}^{4}\,,
\end{equation}
where $\theta(x)\in [0, 2\pi)$ is the azimuthal angular variable and $\rho(x)\equiv |\phi(x)| \in [0, \infty)$ 
is the radial variable, for every space-time point $x$ in Minkowski space $\mathbb{M}^{4}$. The symmetry
group of the theory is $U(1)$. In order to distinguish spontaneous symmetry breaking from the Anderson-Higgs
mechanism, I couple $\phi$ to a $U(1)$-gauge field/connection, $A_{\mu},$ with
$\mu=0,1,2,3$. In studies of superconductivity this gauge field is the electromagnetic vector potential.
A typical action functional for the fields $\phi$ and $A_{\mu}$ has the form
\begin{equation}\label{action}
S(\phi, A_{\mu})=\frac{1}{2}\int d^{4}x\big\{[\partial_{\mu}+ieA_{\mu}(x)]\overline{\phi}(x)\cdot [\partial^{\mu}-ieA^{\mu}(x)] \phi(x) - 
U\big(|\phi(x)|\big)\big\}\,,
\end{equation}
where $e$ is the gauge charge (i.e., twice the elementary electric charge, in the case of superconductivity), 
and $U(|\phi|)$ is the self-interaction potential for the scalar field, which is assumed to
be invariant under the transformation $\phi \mapsto e^{i\theta_0}\, \phi,$ where $e^{i\theta_0}$ belongs to the symmtery 
group $U(1)$. If $\phi$ is written in polar coordinates, $S(\phi, A_{\mu})$ takes the form
\begin{align}\label{action}
S(\phi, A_{\mu}) = \frac{1}{2}\int d^{4}x \big\{ \rho(x)^{2} [\partial_{\mu}\theta(x)-eA_{\mu}(x)]\cdot 
[\partial^{\mu}\theta(x)-eA^{\mu}(x)]
+\partial_{\mu}\rho(x)\cdot \partial^{\mu} \rho(x) - U(\rho(x)) \big\}\,.
\end{align}
I assume that the self-interaction potential $U(\rho)$ is non-negative and has the shape of a 
Mexican hat, with a sharp minimum at $\rho=R>0$, (so that $\rho(x)^{2}\approx R^{2}$).

One distinguishes the following two situations.
\begin{enumerate}
\item{\textit{Continuous symmetry breaking with Goldstone bosons:} We set $e=0$, so that the 
gauge field is absent, and write $S(\phi)$ instead of $S(\phi, A_{\mu})$. Then
\begin{align}
S(\phi)\simeq &S_1(\theta) + S_2(\rho)\,, \quad \text{where}\nonumber\\
S_1(\theta)=& (R^{2}/2)\int d^{4}x\,\, \partial_{\mu} \theta \cdot \partial^{\mu}\theta\,,\label{angular}\\
S_2(\rho)=& \frac{1}{2}\int d^{4}x\, \big\{\partial_{\mu} \rho\cdot \partial^{\mu} \rho - U(\rho)\big\}\,,\label{radial}
\end{align}
and the vacuum expectation value of $\phi$ in a pure vacuum state, $|\emptyset \rangle_{\theta_0}\equiv |\emptyset \rangle$ can be expected to be given by
$$\langle \emptyset |\phi |\emptyset \rangle \approx e^{i\theta_0} R \not= 0, \qquad \text{for some }\, \theta_0 \in [0, 2\pi)\,.$$
The fact that $\big<\emptyset\big|\phi \big| \emptyset\big> \not=0$ and the appearance of the angular 
parameter $\theta_0$ are a signal of spontaneous symmetry breaking.
Expression \eqref{angular} shows that the modes described by the angular field $\theta$ are massless -- 
these are the modes describing the Goldstone bosons that accompany spontaneous symmtery breaking --, 
while the fluctuations of the radial modes described by the field $\rho$ around the minimum of $U$ at 
$\rho=R$ may seem to be massive, assuming that the curvature of $U(\rho)$ at $\rho=R$ is positive. 
Actually, these modes are unstable and decay into pairs of Goldstone bosons. This can be understood
by expanding the selfinteraction potential $U(|\phi|)$ around one of its minima and then noticing
that the fluctuations of the radial field $\rho$ couple to pairs of Goldstone bosons. There
is a rigorous result confirming this claim (see \cite{DN}).

Note that, for a theory in \textit{two} space-time dimensions, the form of the action 
functional \eqref{angular} suggests that the fluctuations of the angular field $\theta$ diverge in 
the infrared, which implies that there is no spontaneous symmtery breaking. This is the content 
of the Mermin-Wagner-Hohenberg (-Coleman) theorem. Continuous symmetries can only be broken in
space-times of dimension $\geq 3$.}
\item{\textit{Anderson-Higgs mechanism:} The gauge charge $e$ does \textit{not} vanish, 
and $A_{\mu}$ is a dynamical gauge field, with a gauge-field action given by
$$S_0(A_\mu) = \frac{1}{4}\int d^{4}x \, F_{\mu \nu}\cdot F^{\mu \nu}\,.$$ 
We may then set
$$\widehat{A}_{\mu}:= A_{\mu} - e^{-1}\partial_{\mu}\theta\,.$$
Obviously, the vector potential $\widehat{A}_{\mu}$ is gauge-equivalent to $A_{\mu}$. We observe 
that the angular field $\theta$ can be eliminated from the theory by a gauge transformation, and the action 
$S(\phi, A_{\mu})$ becomes
\begin{align}
S(\rho, A_{\mu} ) \simeq  \frac{1}{4}\int d^{4}x \big\{ \widehat{F}_{\mu \nu}\cdot \widehat{F}^{\mu \nu} + 2e^{2}R^{2} \widehat{A}_{\mu} \, \widehat{A}^{\mu}\big\} + S_2(\rho)\,, \nonumber
\end{align}
where we have used that $\widehat{F}_{\mu \nu} = F_{\mu \nu}$ (by gauge invariance).

We observe that the gauge field $\widehat{A}$ is \textbf{massive}, with a mass proportional to $(eR)^{2}$, 
and the radial modes described by $\rho$ remain present and are massive; these modes are the ones that
describe the Higgs bosons. (Actually, Higgs bosons are usually unstable particles -- resonances -- that decay
into lighter particles, which the field $\rho$ couples to.)}
\end{enumerate}

Spontaneous symmetry breaking has first been studied, quite a long time ago, for example  in the theory of 
ferromagnetism \cite{Weiss, Heisenberg} and, in particle physics, in the celebrated work of Nambu and 
Jona-Lasinio \cite{NJL}. The fact that the spontaneous breaking of a continuous symmetry is accompanied by
the emergence of massless scalar or pseudo-scalar bosons is the content of the Goldstone theorem
\cite{Goldstone}. 

The arguments leading up to the Higgs mechanism described in item 2 go back, originally, to work 
by St\"uckelberg \cite{Stuckelberg} and Anderson \cite{Anderson}. In particle physics, they were
studied by Englert and Brout; Higgs; Guralnik, Hagen and Kibble \cite{Higgs}. The Higgs mechanism
also works for \textit{non-abelian gauge fields,} which is relevant for the Standard Model of Particle Physics.
The renormalizability of non-abelian gauge fields coupled to Higgs scalars has first been analyzed in
\cite{'tHooft}, and in \cite{LZ-J}.

If the scalar field $\phi$ is coupled to a dynamical gauge field $A_{\mu}$ then it does not make 
sense to speak of spontaneous symmetry breaking. The vacuum expectation value of $\phi$ 
vanishes unless a special gauge is chosen, in which case it can be given an arbitrary value that is 
irrelevant for the physical properties of the theory. A manifestly gauge-invariant formulation of the Higgs
mechanism (not involving any artificial assumptions concerning the vacuum expectation value
of the Higgs scalar $\phi$) has been given in \cite{FMS}. 

The following remarks are limited to the mathematical theory of spontaneous symmetry breaking,
the Goldstone theorem and the Mermin-Wagner-Hohenberg theorem in the context of statistical
mechanics and condensed matter physics. I do not pursue the study of the Higgs mechanism;
but see, e.g., \cite{Weinberg}.

If a Wick rotation to imaginary time is performed on an RQFT one obains what is called a \textit{Euclidean field
theory} (EFT). It has been shown by \textit{Osterwalder} and \textit{Schrader} \cite{OS} (see also \cite{Glaser}) that
from the Euclidean Green functions, also called Schwinger functions, of an EFT satisfying \textit{``Reflection
Positivity''} (which is a manifestation of unitarity at imaginary time) one can reconstruct the vacuum expectation 
values of products of the relativistic (real-time) fields of a corresponding RQFT. EFT's of scalar fields 
can be viewed as systems of statistical mechanics in a state of thermal equilibrium, as first understood 
by \textit{Schwinger, Symanzik} and \textit{Nelson}; see \cite{Simon, GJ} and references given there.
For classical and quantum lattice systems, Reflection Positivity expresses the self-adjointness of the 
transfer matrix of the system in question and gives rise to a variety of crucial bounds on numerous
physical quantities. It plays a central role in many studies of phase transitions and symmetry 
breaking in EFT and in important models of classical and quantum lattice systems, as indicated below.

After a sketch of how the phase transition accompanied by continuous symmetry breaking is
treated in a concrete example of EFT, namely $\lambda |\phi |^{4}$-theory in three dimensions (see Sect.~2), 
I will present a proof of the Goldstone theorem in the formalism of EFT (Sect.~3) and outline consequences 
for $\lambda |\phi |^{4}$-theory in three dimensions (Sect. 3.1). In Sect.~4, I focus attention on systems of 
classical and quantum statistical mechanics in states of thermal equilibrium, and I will summarize some 
important results for lattice theories; see also \cite{Fr} for more detailed surveys.

\section{The phase transition in the $\lambda |\phi |^{4}$-Euclidean field theory in three dimensions}\label{phi4}

The analysis of the phase transition in $\lambda |\phi|^{4}$-theory in three dimensions presented in this
section follows the original work reported in \cite{FSS}.

Henceforth I will always assume that the Wick rotation to purely imaginary time has been made (see \cite{Jost, OS}),
and I will thus only consider Euclidean field theories. Such theories are equivalent to classical Hamiltonian
systems in a state of thermal equilibrium. In this interpretation the Euclidean action functional becomes
the Hamilton functional of the classical system. I will focus my attention on $\lambda |\phi|^{4}$-theory at
imaginary time.

The Euclidean action functional of $\lambda |\phi |^{4}$-theory is given by
\begin{align}\label{action funct}
\begin{split}
S_{\Lambda}(\phi):= &\int_{\Lambda} d^{3}x \big\{\frac{1}{2} \big|\nabla \phi(x)\big|^{2} + U\big(\big|\phi(x) \big|\big) + c.t.\big\}\,,
\quad \text{with}\\
&U(|\phi|)= \frac{\lambda}{4!} \big[|\phi|^{2} - R^{2}\big]^{2} -h\varphi^{1}, \end{split}
\end{align}
where $\phi=(\varphi^{1}, \dots, \varphi^{N})$
is a scalar field with $N=1,2, \dots$ real-valued components (equivalent to a complex scalar field, for $N=2$),
$\Lambda$ is a cube in $\mathbb{R}^{3}$ with sides of length $L$ and with periodic boundary 
conditions imposed at the boundary $\partial \Lambda$, ``$c.t.$'' stands for a logarithmically divergent 
mass counterterm, $\delta m^{2}(\lambda)\, |\phi|^{2}$ (with
$\delta m^{2}(\lambda) \propto \lambda^{2}$), and a divergent (c-number) vacuum-energy counterterm,
$\lambda>0$ is a coupling constant (not renormalized in the three-dimensional theory), 
$R^{2}$ is a real parameter tuning the location of the minima of $U$ in field space and thereby driving
a phase transition accompanied by spontaneous symmetry breaking, and $h\geq 0$ plays the role of a small 
symmetry breaking ``magnetic field'' in the $1$-direction in field space.

We note that, for $h=0$, the $\lambda |\phi|^{4}$-theory with action functional as 
given in \eqref{action funct} has a global $O(N)$ symmetry (with $O(1)= \mathbb{Z}_2$). The theory 
is of obvious physical interest. For $N=1$, it is expected to have the same critical properties as the 
three-dimensional Ising model, thus describing critical phenomena encountered in liquid-gas transitions.
Next, we consider the theory of a complex scalar field equivalent to the theory of a field with $N=2$ real scalar
components.
\begin{itemize}
\item{This theory turns out to describe the ``mean-field limit'' of translation-invariant Bose gases with 
repulsive two-body interactions of very short range (approaching $\delta$-function interactions); see, e.g., 
\cite{FKSS}.  The fact that $\lambda |\phi|^{4}$-theory undergoes a phase transition accompanied 
by spontaneous symmetry breaking, the proof of which is sketched below, is an indication that Bose 
gases with weak repulsive two-body interactions of short range exhibit Bose-Einstein condensation, 
as the temperature is lowered at some fixed positive density -- a conjecture/claim whose 
mathematical proof has remained elusive, so far.}
\item{The critical phenomena observed near the point in parameter space where $\lambda |\phi|^{4}$-theory 
with $N=2$ real scalar fields undergoes a continuous phase transition are expected to be identical to those 
observed for the classical XY-model (rotator) and to be relevant for understanding critical 
phenomena exhibited by superfluid Helium, to mention an example.
Moreover, for $N=3$, the theory is expected to describe critcal phenomena observed in 
three-dimensional ferromagnets near their critical point. See \cite{DGJ} and references given there.}
\end{itemize}
In $d=3$ dimensions, $\lambda |\phi|^{4}$-theory with an action functional given by \eqref{action funct} is a
superrenormalizable theory that can be constructed non-perturbatively by using functional integration.
There are no field-strength- and coupling constant renormalizations. It makes perfect mathematical sense! 
The control of ultraviolet divergences of the theory was first accomplished in \cite{GJ-2}, the 
infinite-volume limit, $L\rightarrow \infty$, was shown to exist in \cite{FO, Park}; 
see also \cite{GJ} and \cite{BFS}. 

In the following we consider the theory in the infinite-volume (``thermodynamic'')
limit $L\rightarrow \infty$, unless stated otherwise. If $\lambda = 0$ (and $L=\infty$) 
then the theory is determined by the free-field \textit{Gaussian measure} of mean 0 and 
covariance $\delta^{ij}(-\Delta)^{-1}$ defined on a space of $\mathbb{R}^{N}$-valued tempered distributions,  
$\mathcal{S}'(\mathbb{R}^{3})^{\times N}$; ($\Delta$ is the three-dimensional Laplacian). The Euclidean one-point 
and two-point Green functions, i.e., the first and second moment of the free-field Gaussian measure, are given by
\begin{equation}\label{free field}
\big< \varphi^{j}(x) \big>_0 \equiv 0, \quad \text{and} \quad \big< \varphi^{i}(x)\, \varphi^{j}(y)\big>_0 =\frac{\delta^{ij}}{4\pi |x-y|}\,.
\end{equation}
The two-point function, $\big< \varphi^{i}(x)\, \varphi^{i}(y)\big>_0$, is given by the Green function of the Laplacian $\Delta$,
for all $i=1,\dots N$.
In the following, $:(\cdot):$ indicates standard Wick ordering with respect to the massless free field 
introduced in \eqref{free field}; for example,
$$:\varphi^{i}(x)\,\varphi^{j}(y): = \varphi^{i}(x)\,\varphi^{j}(y) - \frac{\delta^{ij}}{4\pi |x-y|}\,.$$
We re-write the contribution of the self-interaction potential $U(|\phi|)$ to the action functional $S_{\Lambda}$
as follows.
\begin{align}\label{self-int}
\begin{split}
U_{\Lambda}(R^{2}, h; |\phi|) :=& \int_{\Lambda} d^{3}x\, \big[:U(|\phi(x)|): - \frac{\lambda R^{4}}{4!}\big]\\
=&\int_{\Lambda} d^{3}x \big\{ \lambda :|\phi(x)|^{4}:  -\frac{\lambda R^{2}}{12} :|\phi(x)|^{2}: 
-h\varphi^{1}(x)\big\}\\
=& U_{\Lambda}(0, h; |\phi|) - \frac{\lambda R^{2}}{12} \int_{\Lambda} d^{3}x :|\phi(x)|^{2}:\,.
\end{split}
\end{align}
We then define the ``partition function'' of the theory by
\begin{align}\label{partition funct}
Z_{\Lambda}(R^2, h):= Z_{\Lambda}(0,0)^{-1} \int \mathcal{D}\phi \text{ exp}\big[-\frac{1}{2} \int_{\Lambda} d^{3}x\,
\big| \nabla \phi(x)\big|^{2} + \frac{\lambda R^{2}}{6} :\big|\phi(x)\big|^{2}:+\, c.t.\big]  \, e^{-U_{\Lambda}(0,h;\, |\phi|)}\,, 
\end{align}
and the ``free energy'' as
\begin{equation}\label{free energy}
F(R^{2}, h):= \lim_{L\rightarrow \infty} L^{-3} \ell n Z_{\Lambda}(R^{2}, h)\,.
\end{equation}
The functional integral in \eqref{partition funct} can be given a precise mathematical meaning (see \cite{GJ} 
and references given there). Expectation values with respect to the normalized functional measure
appearing under the integral of the right side of \eqref{partition funct} are denoted by $\big<(\cdot)\big>_{R,h}$. 
The following properties of the function $F(R^{2}, h)$ are known.
\begin{enumerate}
\item[I.]{$F(R^{2}, h)$ is a convex function of $R^{2}$, with
\begin{equation}\label{convexity}
\frac{\lambda}{12}\big< :|\phi(0)|^{2}:\big>_{R, h}= \frac{\partial F(R^{2}, h)}{\partial R^{2}} \nearrow \infty, 
\quad \text{as }\,\,\, R^{2} \rightarrow + \infty\,.
\end{equation}
}
\item[II.]{For fields $\phi$ with N=1, 2, 3 components, the function $F(R^{2}, h)$ is analytic in $h$ away
from the imaginary axis. This is a consequence of a generalized Lee-Yang theorem; see \cite{LY, SG, DN}.}
\end{enumerate}
These statements and the following analysis of the phase transition in $\lambda |\phi|^{4}$-theory in three
dimensions are discussed in detail in \cite{FSS}.

From \eqref{convexity} it follows that, given a finite constant $M^{2}>0$, there exists a 
constant $R^{2}_M< \infty$ such that
\begin{equation}\label{LRO}
\big<:\big|\phi(0)\big|^{2}:\big>_{R,h}\geq M^{2}>0, \quad \text{for }\,\, R^{2}\geq R^{2}_M\,\,
\text{ and all } \,\,\,h\geq 0\,.
\end{equation}
Next we note that
\begin{align}\label{IR bound}
\begin{split}
\big<:\big|\phi(0)\big|^{2}:\big>_{R, h}&= \text{lim}_{y\rightarrow 0} \Big\{\big<\phi(y)\cdot \phi(0)\big>_{R,h} - 
\frac{N}{4\pi |y|}\Big\}\\
&= \text{lim}_{y\rightarrow 0} \Big\{\big< \phi(y)\cdot \phi(0)\big>_{R,h}^{c} + 
\big< \varphi^{1}(0)\big>_{R,h}^{2} - \frac{N}{4\pi |y|}\Big\}\,.
\end{split}
\end{align}
Here $\big<(\cdot)\big>_{R,h}^{c}$ stands for ``connected (truncated) expectation values.'' Since the 
``magnetic field'' $h$ only couples to $\varphi^{1}$, we have that
$\big< \varphi^{i}(0)\big>_{R,h} =0$, for $i=2,\dots, N$, by symmetry, but that
\begin{equation}\label{magnetization}
\big< \varphi^{1}(0)\big>_{R,h} \geq 0, \quad \text{for }\,\, h\geq 0\,.
\end{equation}
In fact, $\big< \varphi^{1}(0)\big>_{R,h}$ is an increasing function of $h$ on the positive half-axis, 
as follows from the fact that the connected two-point function defines a positive quadratic form.

Next, we remind the readers of the so-called K\"allen-Lehmann representation of connected two-point
functions (see, e.g., \cite{Jost} and references given there), which says that
\begin{align}\label{KL}
\begin{split}
\big< \varphi^{i}(y)\cdot \varphi^{i}(0)\big>_{R,h}^{c} = & \int_{0}^{\infty} d\mu^{(i)}(a^2) \big[-\Delta + a^{2}\big]^{-1}(y)\\
\overset{d=3}{=}&  \int_{0}^{\infty} d\mu^{(i)}(a^2) \frac{e^{-a|y|}}{4\pi |y|}\,, \quad \text{for }\,\,i=1,\dots, N\,,
\end{split}
\end{align}
where $d\mu^{(i)}=d\mu^{(j)},$ for $i,j =2,\dots, N$. In $d=3$ dimensions, $\lambda |\phi|^{4}$-theory is a 
canonical field theroy (there is no field strength 
renormalization!). The real-time quantum fields satsify equal-time canonical commutation relations; i.e.,
$$\big[\varphi^{j}(\vec{x}, t), \pi^{k}(\vec{y}, t)\big]= i \delta^{jk} \delta(\vec{x}-\vec{y}), \quad \text{where }\,\, 
\pi^{k} = \dot{\varphi}^{k}\,,$$
and these commutation relations imply that
$$\int_{0}^{\infty} d\mu^{(i)}(a^{2}) =1\,, \quad \forall \,\, i=1,\dots, N\,.$$
It then follows that
$$\big< \phi(y)\cdot \phi(0)\big>_{R,h}^{c} - \frac{N}{4\pi |y|} \leq 0,\quad \forall\,\, y\,,$$
and, using \eqref{LRO} and \eqref{IR bound}, we conclude that
\begin{equation}\label{spontmag}
M^{2}\leq \big<:\big|\phi(0)\big|^{2}:\big>_{R, h} \leq \big< \varphi^{1}(0)\big>_{R, h}^{2},
\end{equation}
for $R^{2}\geq R^{2}_M$, uniformly in $h\geq 0$. Thus, for $R^{2}$ large enough, the theory
predicts ``spontaneous magnetization,'' in the sense that
\begin{equation}\label{spontmag}
\underset{h\searrow 0}{\text{lim }} \big< \varphi^{1}(o)\big>_{R, h} =: M(R) > 0\,.
\end{equation}
Spontaneous magnetization implies that the $O(N)$-symmetry of the action functional of the theory at
$h=0$ -- see \eqref{action funct} -- is spontaneously broken. For one-component fields ($N=1$), 
the existence of a phase transition accompanied by spontaneous breaking of the 
$\phi \rightarrow -\phi$ - symmetry can also be proven in two dimensions; 
but the proof is more involved; (it is based on the so-called Peierls argument -- 
see \cite{GJS} and references given there). For $N\geq 2$, spontaneous magnetization, 
as stated in \eqref{spontmag}, implies that the Fourier transform of the two-point function 
satisfies a lower bound
\begin{equation}\label{LowerB}
\big< \widehat{\varphi}^{i}(k)\, \widehat{\varphi}^{i}(-k)\big>_{R, h=0}\geq \frac{\mathfrak{R}}{|k|^{2}}, 
\quad \forall \,\,i=2,\dots, N\,,
\end{equation}
for some positive constant $\mathfrak{R} = \mathcal{O}(M(R)^{2})$, for small $M(R)$, which 
shows that, in two dimensions and for $N\geq 2$, $M(R)$ must vanish for arbitrary 
values of $R^{2}$, and there is no spontaneous breaking of the $O(N)$-symmetry, because 
$|k|^{-2}$ is not integrable at the origin ($k=0$) in momentum space. The lower bound in 
\eqref{LowerB} can be derived from the analysis in \cite{MW} and can also be extracted from the 
material in Sect.~3. For a more detailed analysis and further references see \cite{Wres}, and vol.~2 
of \cite{BS}.

Next, I draw attention to the fact that, for $N\leq 3,$ if the real part, Re$h$, of the magnetic field $h$ does not vanish then the Green functions
of the theory, such as \mbox{$\big< \phi(x)\cdot \phi(0)\big>_{R, h}$,} are analytic functions of $h$, and 
the connected Green functions decay exponentially in the separation of the field insertion points; in particular 
\mbox{$\big< \phi(x)\cdot \phi(0)\big>^{c}_{R, h}$} decays exponentially in $|x|$ if Re$h \not=0$.
This implies that, for real $h \not= 0$, the physical mass of all modes described by the theory is strictly positive.
For proofs see, e.g., \cite{FR} and references given there. But, for theories with fields of $N\geq 2$ components 
at $h=0$, the Green functions have slow power-law decay if the spontaneous magnetization, $M(R)$, does not vanish. 
In particular, the susceptibility
\begin{equation}\label{susceptibility}
\chi(R, h):= \int_{\mathbb{R}^{3}} d^{3}x \,\big< \varphi^{i}(x)\, \varphi^{i}(0)\big>_{R, h},\quad \text{for }\,\, i\geq 2\,,
\end{equation}
has the property that if the spontaneous magnetization 
$M(R):=\underset{h\searrow 0}{\text{lim}}\big< \varphi^{1}(0)\big>_{R,h}$ is strictly positive then
\begin{equation}\label{divergent}
\chi(R, h) \simeq \frac{M(R)}{h}, \quad \text{ as }\,\,\,h\searrow 0\,.
\end{equation}
Hence, if $R^{2}$ is chosen such that there is spontaneous magnetization, 
in the sense that $M(R)>0$, then, at $h=0+$, the two-point Green functions 
$\big< \varphi^{i}(x)\,\varphi^{i}(0)\big>_{R, 0+}, i\geq 2,$ have non-integrable 
decay in $x$, which shows that there must be massless modes in the theory.
It may be appropriate to sketch the proof of \eqref{divergent}. We consider a change of variables in the 
$(\varphi^{1}, \varphi^{2})$-plane of field space, namely
\begin{align}\label{CofV}
\begin{split}
\varphi^{1}_{\theta} :=& \quad \text{cos}\theta\, \varphi^{1} + \,\text{sin}\theta\, \varphi^{2}\,,\\
\varphi^{2}_{\theta} :=& -\text{sin}\theta\, \varphi^{1} + \,\text{cos}\theta\, \varphi^{2}\,.
\end{split}
\end{align}
where $\theta$ is an $x$-independent angle. In the following, I drop reference to the parameter $R$ (which
will be kept fixed) from my notations and re-write the action functional, $S(\phi)$, of the theory introduced 
in \eqref{action funct} as
$$S(\phi)= \widetilde{S}(\phi) - h \int d^{3}x\, \varphi^{1}(x)\,,$$
where the functional $\widetilde{S}(\phi)$ is manifestly $O(N)$-invariant. Then we have that
\begin{align}\label{Symm}
0=& \big< \varphi^{2}(0)\big>_{h} \nonumber\\
=&\frac{1}{Z(h)} \int \mathcal{D}\phi \,e^{-\widetilde{S}(\phi) + h \int d^{3}x \,\varphi^{1}(x)} \varphi^{2}(0)\nonumber\\
=& \frac{1}{Z(h)} \int \mathcal{D}\phi_{\theta} \,e^{-\widetilde{S}(\phi) + h \int d^{3}x\, \varphi^{1}_{\theta}(x)}\,
\varphi^{2}_{\theta}(0)
\end{align}
identically in $\theta$, where we have used that $\widetilde{S}(\phi_{\theta})= \widetilde{S}(\phi)$, 
by $O(N)$-symmetry. We note that
$$\mathcal{D}\phi_{\theta} = \mathcal{D}\phi\,,$$
because the Jacobian of the change of variables in \eqref{CofV} is unity. By taking the derivative in the variable 
$\theta$ and subsequently setting $\theta=0$ the identity in \eqref{Symm} is seen to imply that
\begin{align}\label{Ward id}
0=& \,\frac{d}{d\theta} \big< \varphi^{2}(0)\big>_{h} \big|_{\theta=0}\nonumber\\
=&\,\frac{1}{Z(h)} \int \mathcal{D}\phi\, e^{-\widetilde{S}(\phi) + h \int d^{3}x \,\varphi^{1}(x)} \big[h\,\int d^{3}x \,
\varphi^{2}(x)\big] \, \varphi^{2}(0)\,  \nonumber\\
&- \frac{1}{Z(h)} \int\mathcal{D}\phi \,e^{-\widetilde{S}(\phi) + h \int d^{3}x\, \varphi^{1}(x)}\, \varphi^{1}(0)\nonumber\\
=& \,h\,\chi(h) - \big< \varphi^{1}(0)\big>_{h}\,.
\end{align}
After re-introducing the parameter $R$ in our notations we see that \eqref{Ward id} and \eqref{spontmag} 
imply that
$$\chi(R, h)= h^{-1} \big< \varphi^{1}(0)\big>_{R, h} \geq \frac{M(R)}{h}\,,$$
which yields \eqref{divergent}. The arguments presented here have been worked out in \cite{Spencer}.

\section{Proof of the Goldstone Theorem for Euclidean Field Theories} 
The following analysis is inspired by a paper on the Goldstone theorem written by the late Kurt Symanzik \cite{Symanzik}.

In comparison with the previous section, we consider a somewhat more general set-up and consider
an RQFT of a scalar field $\phi$ in $d\geq 3$ dimensions with a continuous internal symmetry 
described by a compact Lie group $G$ whose Lie algebra is denoted by $\mathfrak{g}$. A symmetry
transformation generated by an element $X\in \mathfrak{g}$ corresponds to a conserved 
Noether current $j^{\mu}_{X}(x), x=(\vec{x},t) \in \mathbb{M}^{d},$ of the theory; (see, e.g., \cite{BDL} for a 
somewhat abstract treatment of Noether's theorem in RQFT). In the example of 
$\lambda |\phi|^{4}$-theory with an action functional as in \eqref{action funct}, setting $h=0$, the symmetry 
group is $G=O(N)$, and, for $X\equiv X_{ij}:= E_{ij}-E_{ji}$, where $E_{ij}$ is a matrix unit with a $1$ 
in the position $ij$, for $i, j=1, \dots, N$, the corresponding conserved Noether current is given by
\begin{equation}\label{Noether}
j^{\mu}_{X_{ij}}(x) = \big(\partial^{\mu}\varphi^{i}(x)\big) \varphi^{j}(x) - \big(\partial^{\mu}\varphi^{j}(x)\big) \varphi^{i}(x)\,,
\quad x\in \mathbb{M}^{d}\,.
\end{equation}
Returning to the general framework, the field $\phi$ of the theory is assumed to transform under a 
representation $\pi$ of the symmtery group $G$, which we assume not to contain the trivial representation 
of $G$ as a subrepresentation. A generator $X\in \mathfrak{g}$ therefore acts on field space in the 
reperesentation $d\pi$ of $\mathfrak{g}$, i.e., 
$$d\pi(X): \,\, \phi(x) \mapsto d\pi(X)\phi(x) \equiv \phi^{X}(x)\,.$$
Since the currents $j^{\mu}_{X}$ are conserved, i.e., satisfy the continuity equation,
\begin{equation}\label{continuity}
\partial_{\mu} j_{X}^{\mu}(x)=0,
\end{equation}
the charges
\begin{equation}\label{charges}
Q_{X}:= \underset{ t=\text{const.}}{\int} d^{d-1}x\,\, j^{0}_{X} (\vec{x}, t)
\end{equation}
are conserved, and we have that
\begin{equation}\label{action of Q}
\big[Q_{X}, \phi(x)\big]= \phi^{X}(x)\,.
\end{equation}
We now assume that the theory has a vacuum state, denoted by $\big|\emptyset \big> =: \Omega_{\phi_c}$, which is \textbf{not} invariant
under the action of $G$ on the state space of the theory, i.e, breaks the symmetry $G$ spontaneously, with
\begin{equation}\label{vev}
\big<\Omega_{\phi_c}\big| \phi(x)\big|\Omega_{\phi_c}\big> = \phi_c \not= 0\,.
\end{equation}
Let $H_{\phi_c}$ denote the subgroup of $G$ that leaves $\phi_c$ invariant, and let
$\mathfrak{h}_{\phi_c}$ denote its Lie algebra. Then
$$ \phi_{c}^{X}\equiv d\pi(X)\phi_c = 0, \qquad \forall \, X \in \mathfrak{h}_{\phi_c}\,.$$
Since $\pi$ does not contain the trivial represenation of $G$ as a subrepresentation, $\mathfrak{h}_{\phi_c}$ 
is a subspace of $\mathfrak{g}$ of dimension smaller than the dimension of $\mathfrak{g}$. 
For a generator $Y\in \mathfrak{g}$ that belongs to the complement, $\mathfrak{h}_{\phi_c}^{\perp}$, 
of $\mathfrak{h}_{\phi_c}$, we then have that
\begin{equation}\label{vev trsf}
\phi_{c}^{Y}:= \big<\Omega_{\phi_c}\big| \phi^{Y}(x)\big|\Omega_{\phi_c}\big>\not=0\,.
\end{equation}
Combining \eqref{action of Q} with \eqref{vev trsf}, we find that
\begin{align}\label{order param}
\big<\Omega_{\phi_c}\big| \big[Q_{Y}, \phi(0)\big] \big|\Omega_{\phi_c}\big>
=\big<\Omega_{\phi_c}\big| \phi^{Y}(0)\big|\Omega_{\phi_c}\big>
=d\pi(Y) \phi_c \equiv \phi_{c}^{Y}\,,
\end{align}
for an arbitrary $Y \in \mathfrak{h}_{\phi_c}^{\perp}$. On the left side of \eqref{order param} we re-write 
$Q_Y$ using \eqref{charges} and then use a standard formula involving the Hamiltonian $H\geq 0$
of the associated RQFT. This yields
\begin{align}\label{OP - currents}
\begin{split}
\phi_{c}^{Y} = &\,\big<\Omega_{\phi_c}\big| \big[Q_{Y}, \phi(0)\big] \big|\Omega_{\phi_c}\big>\\
  =&\,\underset{t=0}{\int} d^{d-1}x \, \big<\Omega_{\phi_c}\big| \big[j^{0}_{Y}(\vec{x},0), 
  \phi(0)\big] \big|\Omega_{\phi_c}\big>\\
  =& \,\underset{\varepsilon \searrow 0}{\text{lim}} \int d^{d-1}x\, \big<\Omega_{\phi_c}\big| j^{0}_{Y}(\vec{x}, 0)
  e^{-\varepsilon H} \phi(0) - \phi(0) e^{-\varepsilon H} j^{0}_{Y}(\vec{x}, 0) \big|\Omega_{\phi_c}\big>\\
  =&\, \underset{\varepsilon \searrow 0}{\text{lim}} \int d^{d-1}x\, \big<\big[ j^{0}_{Y}(\vec{x}, \varepsilon)
  - j^{0}_{Y}(\vec{x}, -\varepsilon\big]\, \phi(0)\big>\,,
\end{split}
\end{align}
where the integrand on the very right side of these equations is a Euclidean Green- or Schwinger function
of the 0-component of the current $j^{\mu}_Y$ and the field $\phi$ (which, in the example of 
$\lambda |\phi|^{4}$-theory in two or three dimensions, can be expressed by a functional integral).
We define distributions $W^{\mu}_{Y}(x),\, x=(\vec{x},t)\in \mathbb{E}^{d},$ by
\begin{align}\label{W dist}
\begin{split}
W^{\mu}_{Y}(x):=&\, \big< j^{\mu}_{Y}(\vec{x},t)\,\phi(0) \big>\\
     =&\,\begin{cases} \big<\Omega_{\phi_c}\big|j^{\mu}_{Y}(\vec{x}, 0) \,e^{-tH}\,\phi(0)\big|\Omega_{\phi_c}\big>, &
     \,\, \text{ for } \,\, t>0, \\
     \big<\Omega_{\phi_c}\big| \phi(0)\, e^{tH}\,j^{\mu}_{Y}(\vec{x}, 0) \big|\Omega_{\phi_c}\big>, &\,\, \text{ for } \,\, t<0,
     \end{cases}
\end{split}
\end{align}
where we have used that $e^{\pm tH} \big|\Omega_{\phi_c}\big> = \big|\Omega_{\phi_c}\big>$, for arbitrary $t$.
Combining \eqref{order param} through \eqref{W dist}, we find that
\begin{equation}\label{OP}
\phi_c^{Y} = \underset{\varepsilon \searrow 0}{\text{lim}} \int_{S_{\varepsilon}}\,d\sigma_{\mu}(x)\, W^{\mu}_{Y}(x)\not=0\,,
\end{equation}
where $S_{\varepsilon}$ is the union of two oriented planes given by
$$S_{\varepsilon}:= \bigcup_{t= \pm \varepsilon} \big\{ x\in \mathbb{R}^{d}\,\big|\, x= (\vec{x},t), t=\text{const.} \big\}\,,$$
and $d\sigma_{\mu}(\cdot)$ is the oriented surface measure on $S_{\varepsilon}$.
Next, we use that $j^{\mu}_{Y}$ is a conserved current, i.e., that it satisfies the continuity equation \eqref{continuity}
(with $X$ replaced by $Y$), which implies that
\begin{equation}\label{sourceless}
\partial_{\mu}W^{\mu}_{Y}(x) = 0, \quad \text{ for }\,\, x\overset{!}{\not=} 0\,.
\end{equation}
This identity will permit us to apply Gauss' theorem in a useful way. We define two ``half-balls'', 
$B_{r, \varepsilon}^{+}$
and $B_{r, \varepsilon}^{-}$, in $\mathbb{R}^{d}$ with oriented boundaries by
\begin{align}\label{half-balls}
\begin{split}
B_{r, \varepsilon}^{+}:=&\,\big\{ x=(\vec{x},t)\,\big|\, |x| \leq r,  t \geq \varepsilon > 0\big\}\,, \quad \text{with }\,\, |x|:= \sqrt{x^{2}}\,,\\
B_{r, \varepsilon}^{-}:=&\,\big\{ x=(\vec{x},t)\,\big|\, |x| \leq r,  t \leq -\varepsilon < 0\big\}\,.
\end{split}
\end{align}
Since $0\notin B_{r, \varepsilon}^{+}\cup B_{r, \varepsilon}^{-}$, we can apply Gauss' theorem to
$W^{\mu}_{Y}\big|_{B_{r, \varepsilon}^{+}\cup B_{r, \varepsilon}^{-}}$, which then yields
\begin{align}\label{Gauss law}
\begin{split}
0 \not= \phi^{Y}_{c} =&\, \underset{\varepsilon \searrow 0}{\text{lim}} \Big\{ \int_{S_{\varepsilon}} d\sigma_{\mu}(x)\, 
W^{\mu}_{Y}(x)
     + \sum_{\kappa = \pm} \underbrace{\int_{\partial B^{\kappa}_{r, \varepsilon}} \,d\sigma_{\mu}(x)\, W^{\mu}_{Y}(x)\Big\}}_{= 0,\, \text{ by } \eqref{sourceless} \text{ and Gauss' theorem}}\\
     =& \int_{\partial B_r} d\sigma_{\mu}(x)\,W^{\mu}_{Y}(x)\,,
\end{split}
\end{align}
for an arbitrary $r>0$, where$B_r$ is a ball in $\mathbb{R}^{d}$ of radius $r$ centered at the origin,
$x=0$; (the reader is invited to make a drawing to verify this). The second (last) equality in 
\eqref{Gauss law} follows by using Gauss' theorem again. 
Euclidean covariance of $W^{\mu}_{Y}$ implies that
$$ W^{\mu}_{Y}(x) = x^{\mu} f(|x|)\,,$$
for some rotation-invariant function $f$. It then follows from \eqref{Gauss law} that
\begin{equation}\label{Goldstone}
W^{\mu}_{Y}(x) = \frac{\phi^{Y}_{c}}{s_{d-1}} \cdot \frac{x^{\mu}}{|x|^{d}}\,,
\end{equation}
where $s_{d-1}$ is the area of the ($d$-1)-dimensional unit sphere in $d$ dimensions.
The definition of $W^{\mu}_{Y}$ in \eqref{W dist} and Eq.~\eqref{Goldstone} show that the field $\phi$
and the charge density $j^{0}_{Y}$ couple the vacuum state, $\Omega_{\phi_c}$, 
of the theory to a massless scalar one-particle state, which describes a Goldstone boson. Apparently, 
there are at least as many distinct Goldstone bosons in this theory as there are linearly independent 
generators in the complement, $\mathfrak{h}^{\perp}_{\phi_c}$, of $\mathfrak{h}_{\phi_c} \subset \mathfrak{g}$. 
Furthermore, if $\phi_c \not=0$ then the connected two-point function of the field $\phi$ 
has the form
\begin{equation}\label{2-point}
\big< \phi(x)\cdot \phi(0)\big>^{c} = \int_{0}^{\infty} d\mu(a^2) \big[-\Delta+ a^{2}\big]^{-1}(x)\,,
\end{equation}
with
\begin{equation}\label{residue}
d\mu(a^{2}) = \mathfrak{R}\cdot \delta(a^2) \,da^{2} + \text{``regular'' at } a^{2}=0,
\end{equation}
where $\mathfrak{R}$ is the residue of a one-particle pole of zero mass.
 
Well, we have apparently just recovered a proof of the Goldstone theorem in the imaginary-time (Euclidean)
formalism of quantum field theory; (see \cite{Symanzik} for the original arguments).
 
In two dimensions, the residue $\mathfrak{R}$ must vanish if the two-point function of the field $\phi$
is well defined. We thus conculde that $\phi_c$ must vanish, i.e., continuous
symmetries cannot be broken in two dimensions This is the content of the Mermin-Wagner-Hohenberg-Coleman 
theorem; see \cite{Hoh, MW, Coleman}. 

Interesting general results on spontaneous symmetry breaking, Goldstone bosons, and effective Lagrangians
(also for non-relativistic systems of condensed matter) can be found in \cite{Leutwyler}.
 
\subsection{The example of $\lambda |\phi|^{4}$-theory}
 
We conclude this section with some remarks on the $\lambda |\phi|^{4}$-theories with fields $\phi$
of $N=1,2$ or 3 real components. For $N=1$, the symmetry group is discrete, namely $\mathbb{Z}_2$. 
The breaking of discrete groups does not give rise to Goldstone bosons and is possible not only in 
three but also in two dimensions, as shown in \cite{GJS}, where a Peierls-type argument is used to prove
the existence of a phase transition accompanied by the breaking of the $\mathbb{Z}_2$-symmetry.
For $N=2$, the symmetry group is $U(1)$, and symmetry breaking is only possible in dimension 
\mbox{$d\geq 3$.} However, one expects that, in two dimensions, the theory exhibits a Berezinskii-Kosterlitz-Thouless 
transition \cite{BKT} from a massive to a massless phase, but without spontaneous symmetry breaking. 
The existence of this transition has been proven rigorously for the classical XY-model on the 
square lattice and the related Villain model, as well as for some other lattice models, such as the 
two-component Coulomb gas, in \cite{FrSp}; see also \cite{Falco}. But it remains conjectural for the continuum 
$\lambda |\phi|^{4}$- theory of a complex scalar field. For $N\geq 3$, many people expect that, 
in two dimensions, the theory is always massive, i.e., the lowest mass in the theory is strictly positive;
see, e.g., \cite{Polyakov}. 
 
We now turn to the $\lambda |\phi|^{4}$-theories with $N\geq2$ components in three dimensions. For these
theories we have established the existence of a phase transition to a phase with a non-vanishing vacuum
expectation value of the field $\phi$ in a vanishing magnetic field $h=0+$ signaling spontaneous symmetry 
breaking. The analysis presented in the last section shows that if the $O(N)$-symmetry is broken then
there are $N-1$ distinct Goldstone bosons. 
 
 The Lee-Yang theorem implies that, for $N\leq 3$, the connected Green functions of the theory are 
analytic in $h$ and have exponential decay in the separation of their arguments, provided Re$h \not=0$. 
As a corollary, it follows that, for real $h\not=0$, the theory has a strictly positive mass gap $m(h)$, 
i.e., the masses of all excitations in the theory are bounded below by $ m(h)>0$.
 
The results in Sect.~2 yield information on some of the critical exponents of the theories with $N=2, 3$.
We return to Eq.~\eqref{divergent}, which says that the susceptibility $\chi(h)$ of the theory diverges
like $\frac{1}{h}$, as $h\rightarrow 0$, provided the ``spontaneous magnetization'' $M(R)$ 
(see \eqref{spontmag}) does not vanish. A related bound on the mass gap $m(h)$ is plausible,
but remains conjectural. I briefly sketch it here. We have seen in \eqref{2-point} and \eqref{residue}
that if there is spontaneous symmetry breaking then the connected two-point Green function has 
a one-particle pole at zero mass, which, for the example of $\lambda |\phi|^{4}$-theory, with $N\geq 2$
and at $h=0+$, implies that
\begin{equation}\label{particle pole}
\big< \varphi^{i}(x)\cdot \varphi^{i}(0)\big>_{h=0+} = \mathfrak{R}\,[-\Delta]^{-1}(x) + 
\int_{0}^{\infty} d\mu_{reg}(a^2) [-\Delta + a^{2}]^{-1}(x)\,, \quad \text{ for }\,\,i \geq 2\,,
\end{equation}
where the residue $\mathfrak{R}$ of the one-particle pole at zero mass is strictly positive in
the symmetry-broken phase, and the measure $d\mu_{reg}(a^2)$ is regular at $a^{2}=0$. 
(It is assumed here that the parameter $R^{2}$ is large enough such that there is spontaneous
symmetry breaking; but we do not display the $R$-dependence explicitly.)
If $h>0$ this representation takes the form
\begin{equation}\label{massive particle}
\big< \varphi^{i}(x)\cdot \phi^{i}(0)\big>_{h} = \mathfrak{R}(h)\,[-\Delta+m^{2}(h)]^{-1}(x) 
+ \int_{m^{2}(h)}^{\infty} d\widetilde{\mu}(a^2) [-\Delta + a^{2}]^{-1}(x)\,,\quad i\geq 2\,,
\end{equation}
where the physical mass, i.e., the inverse correlation length, $m(h)$, is strictly positive , 
as follows from the Lee-Yang theorem, and $d\widetilde{\mu}(a^2)$ 
is regular at $a^{2}=m^{2}(h)$. It is plausible to expect that $\mathfrak{R}(h)\approx \mathfrak{R}$ 
is strictly positive, for $h>0$ small enough, i.e.,
\begin{equation}\label{LB}
\mathfrak{R}(h)\geq \mathfrak{R}_{*} >0, \qquad \text{for }\,\, h>0\,\, \text{ small enough}\,.
\end{equation}
Using convergent expansions, this expectation is known to be true if $h$ is real and $|h|$ 
is large enough. However, the strict positivity of $\mathfrak{R}(h)$ for small, but positive 
values of $h$ is a conjecture that has not been proven, so far. Assuming that it holds true, 
we conclude from \eqref{massive particle} and \eqref{LB} that, for $i\geq 2$,
\begin{equation}\label{bound on mass}
\chi(h)=\int d^{3}x \big< \varphi^{i}(x)\cdot \phi^{i}(0)\big>_{h} \geq \frac{\mathfrak{R}_{*}}{m^{2}(h)}, \quad \text{ with }\,\, \mathfrak{R}_{*} >0\,.
\end{equation}
Combining this lower bound with expression \eqref{divergent} for the susceptibiliity $\chi(h)$, we find that
the mass gap $m(h)$ of the theory satisfies the lower bound (predicted by heuristic arguments)
\begin{equation}\label{LB-mass}
m(h)\geq \text{const.} \sqrt{h}\,,\qquad \text{as }\,\, h\searrow 0\,.
\end{equation}
One can rigorously prove a related upper bound on $m(h)$ as follows. From the K\"allen-Lehmann representation, 
$$\big< \varphi^{i}(y)\cdot \varphi^{i}(0)\big>_{\rho,h}^{c} = \int_{m^{2}(h)}^{\infty} d\mu^{(i)}(a^2) \big[-\Delta + a^{2}\big]^{-1}(y)\,, \quad \, i\geq 2\,,$$
see \eqref{KL}, the fact that $\int d\mu^{(i)}(a^2) =1$, and \eqref{divergent} we derive that
$$\frac{M(R)}{h} \leq \chi(h) \leq \frac{1}{m(h)^2}\,$$
 hence if there is spontaneous magnetization ($M(R)>0$) then, for positive $h$,
 \begin{equation}\label{crit expo}
 m(h) \leq \text{const.}\sqrt{h}, \qquad \text{as }\,\, h\searrow 0.
 \end{equation}
  It is evident that whatever is written here for $h>0$ also holds for $h<0$ (by replacing $\varphi^{1}$
 by $-\varphi^{1}$).
 
 Equations \eqref{divergent}, \eqref{LB-mass} and \eqref{crit expo} are results on critical exponents 
 describing the behavior of the susceptibility and of the mass gap, as $h\searrow 0$, assuming that the
 spontaneous magnetization is strictly positive. It would be of considerable interest to understand the
 behavior of the spontaneous magnetization, $M(R)$, defined in \eqref{spontmag}, as the
 parameter $R^{2}$ approaches the phase-transition point, i.e., $R^{2} \searrow R^{2}_{c}$, with
 $M(R) = 0$, for $R^{2} < R^{2}_{c}$.
 One expects that $M(R)\searrow 0$ continuously, as $R^{2} \searrow R^{2}_c$.
 But this has only been proven rigorously for the three-dimensional Ising model (corresponding to $N=1$), 
 with $M(R) \geq \mathcal{O}\big([R^{2}-R^{2}_{c}]^{-1/2}\big)$; see \cite{A-DC-S}. 
 
One may ask how, in zero magnetic field, $h=0$, the suceptibility, $\chi(R,h=0)$, and the mass gap, 
$m(R,h=0)$,  behave when $R^{2} \nearrow R^{2}_c$. A correlation inequality for the connected four-point 
(Ursell) function due to Lebowitz \cite{Lebowitz} implies that, for one-component real fields ($N=1$),
$\chi(R, 0)$ diverges at least as rapidly as $\big(R^{2}_c - R^{2}\big)^{-1}$, and the mass gap 
$m(R, 0)$ tends to 0 continuously, as $R^{2}\nearrow R^{2}_c$; see \cite{GJ_crit, McBryan}. 
(These results are also known to hold for complex scalar fields ($N=2$), but not for $N\geq 3$.)
 
 The reader may wonder why the $\lambda |\phi|^{4}$-theory in four (or more) dimensions has not been
 considered so far. The reason is that if all ultraviolet cutoffs are removed this theory is expected
 to be equivalent to a (generalized) free field. Assuming that one regularizes the theory by putting it
 on a hypercubic lattice, this result has been proven for the theories in dimension $d>4$, with fields of
 $N=1$ or 2 components, in \cite{Aizenman, Fr_82}. In four dimensions, this result is only known for the 
 scaling limit of the Ising model and for $\lambda \phi^{4}_4$-theory of a one-component real scalar field, 
 the very beautiful, subtle proof appearing in \cite{A-DC}. 
 Partial results for $\lambda |\phi|^{4}_4$-theory of a complex scalar field in four dimensions have
 been reported in \cite{Fr_82}; but these results remain incomplete. For $N\geq 3$, there are
 no rigorous results known, yet.
 
 These remarks motivate us to consider classical and quantum lattice systems, which do not have an ultraviolet 
 (short-distance) problem. I will provide a guide to the literature in the next section, but won't discuss any
 details.
 
 \section{Phase transitions and symmetry breaking in classical and quantum lattice systems}
For applications in statistical mechanics and condensed matter physics, classical and quantum lattice systems
are usually more relevant than continuum field theories, although the latter tend to describe critical phenomena 
of lattice systems near continuous phase transitions. 

We limit our review to sketching some results for classical or quantum spin systems on a lattice 
$\mathbb{Z}^{d}$, with $d=2,3,4$. 
The Hamiltonian of such systems is given by
\begin{equation}\label{Hamiltonian}
H\equiv H(\{\vec{S}\}):= -\sum_{x,y \in \mathbb{Z}^{d}} \big\{J_{\perp}(x-y) \big[S^{1}_x\cdot S^{1}_y + 
S^{2}_x\cdot S^{2}_y\big] +J_{\Vert}(x-y) S^{3}_x
\cdot S^{3}_y\big\} - h \sum_{x\in \mathbb{Z}^{d}} S^{3}_x \,,
\end{equation}
where $J_{\perp}(x-y)$ and $J_{\Vert}(x-y)$ are exchange couplings (see \cite{Heisenberg}), 
$h$ is proportional to the strength of an external magnetic field in the 3-direction, and 
$\vec{S}_x=\big(S^{1}_x, S^{2}_x, S^{3}_x\big)$ is either a classical spin in $\mathbb{R}^{3}$
or a quantum spin with spin quantum number $s=\frac{1}{2}, 1, \dots$ 
\begin{enumerate}
\item[(i)]{If $J_{\Vert}(x-y) \gg |J_{\perp}(x-y)|, \,\forall\, x, y,$ this Hamiltonian describes uniaxial 
magnetism. For classical spins or quantum spins with $s=\frac{1}{2}$, the model behaves like an 
Ising model with $J_{\perp}\equiv 0$. 
Phase transitions only occur when $h=0$, in which case the model has a $\mathbb{Z}_2$-symmetry. 
Rigorous results for quantum spins can be found in \cite{Ginibre, Kennedy} and references given there.}
\item[(ii)]{If $J_{\perp}(x-y) \gg |J_{\Vert}(x-y)|,\, \forall \, x, y$, the model describes a classical or 
quantum XY (rotator) model, which has a continuous symmetry $U(1)$.}
\item[(iii)]{If $J_{\perp}\equiv J_{\Vert}=:J$ the model describes an isotropic magnet and, 
for $h=0$, has a continuous symmetry given by $SO(3)$ or $\widetilde{SO}(3)= SU(2)$.}
\end{enumerate}
Let us first consider the classical models. They have been studied in \cite{FSS} and, using totally different
methods, in \cite{GSp}; and references given there. We briefly review the results and methods used in \cite{FSS}, 
which are inspired by properties of RQFT's and EFT's, in particular by the K\"allen-Lehmann representation 
\eqref{KL}. 

For classical spin systems, the number of components of the spin $\vec{S}$ is irrelevant; we may
replace $(S^1, S^2)$ by $(S^1, \dots, S^{N-1})$ and $S^3$ by $S^{N}$, $N=1,2,3,\dots$
We begin by explaining what it means for exchange couplings $J$ to be ``reflection positive.''
Let $\Pi$ be a hyperplane in $\mathbb{R}^{d}$ of co-dimension 1 that lies in between two lattice
planes of $\mathbb{Z}^{d}$ (viewed as a subset of $\mathbb{R}^{d}$), and let $\theta$ denote reflection
at $\Pi$. Then the two sites $x\in \mathbb{Z}^{d}$ and $\theta x \in \mathbb{Z}^{d}$ lie symmetrically
on two different sides of the hyperplane $\Pi$. We define $\mathbb{Z}^{d}_{+}$ and $\mathbb{Z}^{d}_{-}$ to be
the two half-lattices on the two different sides of the hyperplane $\Pi$. Let $J(x-y), x, y \in \mathbb{Z}^{d},$ be
exchange couplings. 

\textit{\underline{Definition}:} $J$ is said to be ``reflection positive'' iff
\begin{equation}\label{RP}
\sum_{x, y \in \mathbb{Z}^{d}_{+}} J(x-\theta y) z_x \overline{z}_{y} \geq 0\,,
\end{equation}
for an arbitrary complex-valued function $z$ on  the half-lattice $\mathbb{Z}^{d}_{+}$. \hspace{0.6cm} {\tiny{$\square$}}

The equilibrium state of a classical spin system with a Hamiltonian given by \eqref{Hamiltonian} at inverse
temperature $\beta$ is given by the formal probability measure
\begin{equation}\label{eq state}
dP(\{\vec{S}\}):= \frac{1}{Z_{\beta}} e^{-\beta H(\{\vec{S}\})} \prod_{x \in \mathbb{Z}^{d}} d\rho(\vec{S}_x)\,.
\end{equation}
where $d\rho(\vec{S})$ is a finite measure on $\mathbb{R}^{N}$, for example
\begin{align}\label{single spin}
d\rho(\vec{S})=\begin{cases} \text{exp}\big(-\frac{\lambda}{4!} \big[\big|\vec{S}\big|^{2}-R^{2}\big]^{2}\big)\,d\vec{S}\,, 
\quad \text{or}\\
\delta(|\vec{S}|^{2} - R^{2})\,d\vec{S}\,,
\end{cases}
\end{align}
with $d\vec{S}$ the Lebesgue measure on $\mathbb{R}^{N}, N=1,2,3,\dots$, and $Z_{\beta}$ the partition function, 
which is chosen such that $\int dP(\{\vec{S}\}) = 1$. Since we have not normalized
the exchange couplings, we may henceforth set $\beta =1$ and omit it. Expectations of products
of components of spins with respect to the measure $dP(\{\vec{S}\})$ are denoted by 
$\big<(\cdot)\big>_{J_{\perp}, J_{\Vert},h} \equiv \big<(\cdot)\big>$.

It turns out that, for classical spin systems with a Hamiltonian given by \eqref{Hamiltonian} and reflection-positive 
exchange couplings $J_{\perp}$ and $J_{\Vert}$ satisfying \eqref{RP}, there is an analogue of the K\"allen-Lehmann
representation in the form of upper bounds, so-called Infrared Bounds, on the Fourier transforms of connected 
spin-spin correlation functions: for $k\not=0$,
\begin{align}\label{IR bounds}
\begin{split}
0< \big< \widehat{S}^{i}(k)\,& \widehat{S}^{i}(-k)\big> \leq 
\frac{1}{\widehat{J}_{\perp}(0)-\widehat{J}_{\perp}(k)}\,, \quad \text{ for }\,\, i=1,\dots, N-1,\\
0&< \big< \widehat{S}^{N}(k)\, \widehat{S}^{N}(-k)\big> \leq 
\frac{1}{\widehat{J}_{\Vert}(0)-\widehat{J}_{\Vert}(k)}\,,
\end{split}
\end{align}
where $\widehat{(\cdot)}$ denotes Fourier transformation, that is $\widehat{S}^{i}(k)$ is the Fourier 
transform of $S^{i}_x$, for $i=1, \dots, N$, and $k\in B$, where $B$ is the Brillouin zone (i.e., a 
$d$-dimensional torus, for the lattice $\mathbb{Z}^{d}$).
At the origin, $k=0$, in quasi-momentum space, these two-point functions may include $\delta$-function
contributions, $M_{i}^{2}\,\delta_{0}(k)\,d^{d}k$, where the constants $M_{i}^{2}$ manifest long-range order (LRO), 
for $i=1,2,\dots,N$. In an extremal equilibrium state, the absolute value of the spontaneous magnetization 
is then given by $\mathcal{M}=\sqrt{\sum_{i=1}^{N} M_{i}^{2}}$. The bounds in \eqref{IR bounds} 
have been established rigorously in \cite{FSS} and \cite{FILS}.

We now show that, in $d\geq 3$ dimensions and for exchange couplings $J_{\perp}$ and $J_{\Vert}$
of short range, the quantity $\mathcal{M}$ must be strictly positive, provided $R$ is large enough. 
First we note that, for $d\rho$ as in \eqref{single spin},
\begin{align}\label{sum rule}
\sum_{i=1}^{N} \int_{B} d^{d}k\, \big<\widehat{S}^{i}(k)\,\widehat{S}^{i}(-k)\big> = \big<\vec{S}_{0}\cdot \vec{S}_{0}\big> = c \cdot R^{2}\,,
\end{align}
for some strictly positive constant $c$. (This is obvious for the choice 
$d\rho(\vec{S})=\delta(|\vec{S}|^{2} - R^{2})\, d\vec{S}$, in which case $c=1$.)
But, by \eqref{IR bounds}, the left side is bounded from above by
\begin{equation}\label{upper bound}
\big<\vec{S}_{0}\cdot \vec{S}_{0}\big> \leq \int_{B} d^{d}k \Big[\frac{N-1}{\widehat{J}_{\perp}(0)-\widehat{J}_{\perp}(k)}
+\frac{1}{\widehat{J}_{\Vert}(0)-\widehat{J}_{\Vert}(k)}\Big] + \mathcal{M}^{2}.
\end{equation}
If the exchange couplings $J_{\perp}$ and $J_{\Vert}$ are of short range then
\begin{align}\label{IR bounds-2}
\big[\widehat{J}_{\alpha}(0) - \widehat{J}_{\alpha}(k)\big]^{-1} \leq J_{\alpha}^{-1}|k|^{-2}\,,
\end{align}
for some positive constants $J_{\alpha}$, with $\alpha \,= \,\perp \text{ or }\, \Vert$. Taking 
$J:=\text{min}(J_{\perp}, J_{\Vert})$ we find that
\begin{equation}\label{54}
c\cdot R^{2}= \big<\vec{S}_{0}\cdot \vec{S}_{0}\big> \leq \frac{N}{J} I_d + \mathcal{M}^{2}\,.
\end{equation}
where $I_d=\int_{B}\frac{d^{d}k}{k^{2}}$. Obviously, $I_d$ diverges
in $d=1$ and 2 dimensions, in which case \eqref{54} is useless. But $I_d$ is finite for $d\geq 3$, 
and it then follows that 
$$\mathcal{M}^{2}\geq c\cdot R^{2}- \frac{N}{J} I_d>0\,, \quad \text{for }\,\, R^{2}\,\, \text{big enough}\,,$$
i.e., there is spontaneous magnetization and symmetry breaking if $R^{2}$ is big enough.

For the Ising model ($N=1$) and the classcial XY model ($N=2$), these results on the existence
of phase transitions accompanied by symmetry breaking can be extended to a large class of
not necessarily translation-invariant and not necessarily reflection-positive exchange couplings
by using what one calls ``correlation inequalities.'' The relevant inequalities can be found in
\cite{Messager}.

One expects that if $J_{\Vert} > J_{\perp}$ then the spontaneous magnetization is parallel to the $N$-axis
(with $N=3$ for the Hamiltonian in \eqref{Hamiltonian}). In this case, the discrete symmetry 
\mbox{$S^{N} \rightarrow -S^{N}$} is broken; for results in this direction see \cite{Kennedy} and
references given there. But if $J_{\perp} > J_{\Vert}$ then the spontaneous magnetization 
is expected to lie in the plane $\{\vec{S}\,|\,S^{N}=0\}$, and the continuous symmetry $O(N-1)$ of the 
Hamiltonian is broken, which, for $N\geq 3$, is accompanied by the emergence of $N-2$ 
Goldstone bosons. If $J_{\Vert}=J_{\perp}\equiv J$ then the model exhibits an $O(N)$-symmetry, 
and the spontaneous magnetization can have an arbitrary direction in $\mathbb{R}^{N}$, which 
can be chosen by introducing a small symmetry breaking term, $-\vec{h}\cdot \sum_{x} \vec{S}_x$, 
in the Hamiltonian and letting $|\vec{h}|$ tend to 0 eventually. All these predictions can be shown 
to hold rigorously for $N=1, 2, 3$, under sutiable assumptions on the parameters.

It may be appropriate to sketch a heuristic idea that originally inspired a stronger (exponential) 
version of the Infrared Bounds in \eqref{IR bounds},  and which could then be extended to quantum
lattice systems. We consider a canonical quantum field theory of a real
scalar field $\varphi$ on $d$ space-time dimensions with a Hamiltonian, denoted by $H$, whose
energy spectrum is bounded from below by 0.
Such a theory (if it existed) would have the following properties:
\begin{align}\label{CCR}
\begin{split}
\dot{\varphi}(\vec{x}, t) = -i \big[H, \varphi(\vec{x},t)\big] &=: \pi(\vec{x}, t)\,, \quad \text{with}\\
\big[\varphi(\vec{x}, t), \pi(\vec{y}, t)\big] &= i\,\delta(\vec{x}-\vec{y})\,,
\end{split}
\end{align}
where $t$ denotes time and $\vec{x}$ is a point in space.
Let $f$ and $g$ be real-valued test functions on $\mathbb{R}^{d-1}$, and let 
\begin{align}\label{smearing}
&\varphi(f, t):= \int_{\mathbb{R}^{d-1}} d\vec{x} \,\,\varphi(\vec{x},t)\, f(\vec{x})\,,\\
&\pi(g, t):= \int_{\mathbb{R}^{d-1}} d\vec{x}\, \,\pi(\vec{x},t)\, g(\vec{x})\,, \quad\, \text{with}\\
&\big[\varphi(f,t), \pi(g,t)\big]=i \int_{\mathbb{R}^{d-1}} d\vec{x}\,\, f(\vec{x})\, g(\vec{x}) \equiv i\,\big<f, g\big>\,.
\end{align}
Since $\varphi$ is a real field, we may expect that, at all times $t$, $\varphi(f, t)$ is a selfadjoint operator on the 
Hilbert space, $\mathcal{H}$, of the theory. From our assumption that $H\geq0$ it follows that 
$U\,H\,U^{*} \geq 0,$ for an arbitrary unitary operator $U$ on $\mathcal{H}$, and we then find that
\begin{equation}\label{pi-bound}
0 \leq e^{i\varphi(f,t)}\,H\,e^{-i\varphi(f,t)} = H + \pi(f, t) + \frac{1}{2}\Vert f\Vert_{2}^{2}\cdot\mathbf{1}\,,
\end{equation}
where $\Vert f\Vert_{2}^{2} := \big< f, f\big>$. Using a Feynman-Kac formula, one can translate 
this bound to one that holds for a corresponding EFT at imaginary time. We denote the Euclidean field 
corresponding to the quantum field $\varphi$ by $\phi$, as before, and expectations with respect to the 
functional measure describing the EFT corresponding to the QFT are denoted by $\big<(\cdot)\big>$. 
For a real-valued test function $h$ on $\mathbb{R}^{d}$, we define $\phi(h):=\int d^{d}x\,\, \phi(x)\,h(x)$.
Then \eqref{pi-bound} implies that
$$\big< e^{\frac{\partial \phi}{\partial t}(h)}\big> \leq e^{\frac{1}{2}\Vert h \Vert_{2}^{2}}\,.$$
By Euclidean invariance, it then follows that
\begin{equation}\label{expo IR}
\big< \text{exp}\big[\underline{\nabla}\phi(\underline{h})\big> \leq e^{\frac{1}{2} \Vert \underline{h} \Vert_{2}^{2}}\,,
\end{equation}
where $\underline{h}=(h_1, \dots, h_d)$ is a $d$-tuple of test functions. 
This bound readily implies that, in momentum space and for $k\not=0$,
$$\big< \widehat{\phi}(k)\, \widehat{\phi}(-k)\big> \leq \frac{1}{|k|^{2}}\,,$$
which is an Infrared Bound similar to \eqref{IR bounds} in the special case where $N=1$ and where
$J_{\Vert}$ is a nearest-neighbor coupling of strength 1. Obviously, one can extend these arguments to
canonical theories of $N$-component real fields, $\vec{\varphi}=(\varphi^1, \dots, \varphi^N), N\geq 2$. 
(Of course, the bound on $\big< \widehat{\phi}(k)\, \widehat{\phi}(-k)\big>$ also follows 
from the K\"allen-Lehmann representation \eqref{KL}.)

For classical lattics spin systems with nearest-neighbor exchange couplings, bounds of the form of 
\eqref{expo IR} have first been proven in \cite{FSS}. They were subsequently extended, mutatis mutandis, 
to classical models with general reflection-positive exchange coupling, including long-range couplings, 
in \cite{FILS} and to quantum spin systems in \cite{DLS}.

A much more ambitious, more robust analysis of classical lattice spin systems with phase transitions
accompanied by the spontaneous breaking of continuous $O(N)$-symmetries has been undertaken
in \cite{Balaban}, using mathematically controlled renormalization group methods. (See also 
\cite{Bauerschmidt} for a somewhat related, but simpler analysis of a different model.)

A new, quite surprising and original approach to proving phase transitions accompanied by spontaneous 
symmetry breaking has been proposed in \cite{GSp}. The advantage of their method is that it also applies 
to classical models that are neither translation-invariant nor do they satisfy Reflection Positivity. The
price to pay is that the models describe disordered systems and have random exchange 
couplings. (For the Ising- and the classical XY model, one can, however, use the correlation inequalities 
in \cite{Messager} to extend their results to the usual choice of exchange couplings; see \cite{GSp}.)

It turns out that Infrared Bounds analogous to those in \eqref{IR bounds} -- but for so-called Duhamel
two-point correlations -- can also be proven for quantum lattice spin systems satisfying Reflection
Positivity, as discovered in \cite{DLS}. It turns out that, among the quantum Heisenberg models, only
the anti-ferromagnets satisfy Reflection Positivity, but not the ferromagnets. For anti-ferromagnets, the Infrared 
Bounds can be used to establish phase transitions accompanied by spontaneous symmetry breaking
and the emergence of Goldstone modes (magnons); see \cite{DLS}. More general results
on phase transitions and spontaneous symmetry breaking for a large family of quantum lattice systems 
satisfying Reflection Positivity have been established in \cite{FILS, Albert}; see also \cite{Fr} for reviews.
I cannot go into any details about these results. For the isotropic quantum Heisenberg ferromagnet on the 
complete graph, the existence of a phase transition accompanied by the spontaneous breaking of the 
$SU(2)$-symmetry has been proven in \cite{BFUe}, using exact calculations that do not rely on Reflection 
Positivity; but rigorous results concerning phase transitions in quantum Heisenberg ferromagnets with 
continuous internal symmetries on finite-dimensional lattices are not known, so far.

I conclude this review with a few comments on the Ising model and the classical XY model. By Kramers-Wannier
duality, these models are equivalent to gases of ``topological defects.'' For the Ising model, these defects are
domain walls (also called Peierls contours) separating domains with opposite spin orientation. 
For the XY model the defects are (co-dimension-2) vortices. It turns out that, in dimension $d\geq 2$ and
at low temperatures, these defect gases are dilute. This feature enables one to analyze them by using 
mathematically controlled versions of energy-entropy arguments underpinned by ``multi-scale analysis.'' 
For the Ising model, this strategy is implemented in what is commonly called a ``Peierls argument,'' 
in honor of its inventor. For the XY model, the analysis is considerably more subtle, due to
the presence of massless Goldstone modes at low temperatures. In particular, the analysis 
of the low-temperature phase of the two-dimensional classical XY model, viewed as a dilute gas of point
vortices with long-range (Coulomb) interactions, is really rather involved. An interesting, fairly precise upper
bound on spin-spin correlcations in the two-dimensional XY model at low temperatures has been 
proven in \cite{McB-Sp}. Techniques (involving complex translations in the space of spin configurations) 
developed in this paper have provided some of the tools used in \cite{FrSp}, where, using rather heavy
multi-scale analysis, the existence of the Kosterlitz-Thouless transition for the two-dimensional classical 
XY model and other related models has been established rigorously; see also \cite{Falco}. New, simpler
proofs have recently appeared in \cite{Lammers} (and references given there).

One of the key results established in \cite{McB-Sp, FrSp} is the following bound on the spin-spin correlation.
For ferromagnetic exchange couplings $J > J_c$, where $J_c$ is the coupling at which the Kosterlitz-Thouless 
transition occurs,
\begin{align}\label{spin-spin}
\begin{split}
\big<\vec{S}_0\cdot \vec{S}_x\big>_{J, h=0} \simeq \mathcal{O}\big(\big[|x|&+1\big]^{-\{1/2\pi \varepsilon(J)\}}\big)\,, \quad \text{with}\\
0<\,\,\varepsilon(J) \rightarrow J&,\quad \text{ as }\,\,\, J\rightarrow \infty\,.
\end{split}
\end{align}
Here $\vec{S}_x=(\text{cos }\theta_x, \text{sin }\theta_x), \, \theta_x \in [0, 2\pi),$ is the planar spin variable
of the classical XY model at site $x\in \mathbb{Z}^{2}$, and the external magnetic field $h$ vanishes, $h=0$. 
The Lee-Yang theorem, which holds for the classical XY model,
implies that the susceptibility 
$$\chi(J, h)=\sum_{x\in \mathbb{Z}^{2}} \big<\vec{S}_0\cdot \vec{S}_x\big>_{J, h}$$ 
is finite if $h>0$. By \eqref{spin-spin}, $\chi(J,h)$ diverges, as $h\searrow 0$, provided $J$ is large enough.
The magnetization, $M(J, h)$, is non-zero, for $h>0$, but approaches 0, as $h\searrow 0$,
by the Mermin-Wagner-Hohenberg theorem \cite{Hoh, MW}. Using that
$$\chi(J,h) = \frac{M(J,h)}{h}\,,$$
see \eqref{divergent}, we conclude that, for $J$ large enough,
\begin{align}\label{crit-exp}
\begin{split}
M(J,h) \rightarrow 0, &\quad \text{ but }\,\,\frac{M(J,h)}{h} \,\, \text{ diverges,\,\, as }\,\, h\searrow 0\,,\,\, \text{ and}\\
&\chi(J,h) = \mathfrak{o}(h^{-1})\,, \,\, \text{ as }\,\,h\searrow 0\,,\\
\end{split}
\end{align}
while, for $J<J_c$, where the spin-spin correlation decays exponentially, as $|x|\rightarrow \infty$, and
the susceptibility remains finite, as $h\searrow 0$, we have that $M(J,h) \propto h,$ as $h\searrow 0$.

For the three-dimensional classical XY model, the analysis is simpler, because, in $d=3$, the vortices 
are one-dimensional objects and form loops. At high temperatures, these vortex loops condense; but,
at low temperatures, they form a dilute gas. They have long-range interactions mediated by massless 
Goldstone modes. Somewhat surprizingly, these long-range interactions do not cause any major 
complications in the mathematical analysis of the low-temperature phase of the model. In fact, the vortex-loop 
representation has been used to rigorously prove that, at low enough temperatures, the three-dimensional 
classical XY model has spontaneous magnetization and that its connected spin-spin correlations decay 
very slowly, namley like $r^{-1}$, where $r$ is the distance between the spin insertion points. 
It is easy to show that, at high temperatures, the spin-spin correlations decay exponentially fast. 
These findings establish again the existence of a phase transition accompanied by the spontaneous 
breaking of a $U(1)$-symmetry and the emergence of Goldstone bosons; see \cite{F-Sp}. 
Translation invariance of the Hamiltonian is not essential in this method.

Ideas somewhat related to those in \cite{F-Sp} have earlier been used in \cite{Guth} in an 
analysis of the deconfining transition in the four-dimensional $U(1)$-lattice gauge theory; 
see also \cite{F&S}.\\

\noindent
\underline{Some Conclusions and Open Problems}:\\

I hope this short review convinces the reader that many non-trivial facts about phase transitions accompanied
by spontaneous symmetry breaking and about Goldstone bosons of considerable interest in condensed
matter and particle physics are now known in the form of mathematically precise results. Work on this 
subject has given rise to many important insights into properties of physical systems and to the 
development of numerous interesting techniques of analysis that can also 
be used in other contexts. Ideas from statistical and condensed matter physics have been fruitfully 
imported into particle physics and quantum field theory, and conversely. Without this cross-fertilization, 
many significant results would presumably not have been found, yet. 

I expect that I have quoted sufficiently much of the relevant literature to enable the reader to access without
effort the nitty-gritty of the topics discussed in this review. (I offer my apologies to all colleagues whose 
work I should have quoted, but did not.)

Despite all the progress that has been made during more than half a century, many important questions remain
open, at least if one insists on adhering to the standards of mathematical physics. Among the most important 
such questions are the following ones.
\begin{enumerate}
\item{Find a continuum model with a translation-invariant Hamiltonian for which one can prove 
that the ground-state and the low-temperature equilibrium states break translation invariance spontaneously
and exhibit crystalline ordering.}
\item{Show that translation-invariant interacting Bose gases with physically realistic repulsive two-body 
interactions exhibit Bose-Einstein condensation as the temperature is lowered at a fixed, positive density.}
\item{Show that the isotropic quantum Heisenberg ferromagnet in $d\geq 3$ dimensions undergoes 
a phase transition accompanied by the spontaneous breaking of its $SU(2)$-symmetry.}
\item{Show that the quantum-mechanical XY model and the $\lambda |\phi|^{4}$-theory in two 
dimensions undergo a Kosterlitz-Thouless transition.}
\item{Prove rigorously that the classical Heisenberg model and the classical $N$-vector models, with
$N > 3$, have a strictly finite correlation length at all positive temperatures.}
\item{Extend the results in \cite{A-DC-S, A-DC} concerning the magnetization in the three-dimensional 
Ising model, and the Gaussian nature of the scaling limit of the four-dimensional Ising- and 
$\lambda \phi^{4}$-model ($N=1$), respectively, to two-component models, in particular 
to the classical XY model.}
\end{enumerate}

\underline{Acknowledgements}. This review could not have been written without my collaborations, mostly many
years ago, with E.~H.~Lieb, B.~Simon and T.~Spencer. They have taught me very many important 
things about statistical mechanics and quantum field theory. My collaboration and my countless 
discussions, also in more recent times, with T.~Spencer count among the highlights in my scientific 
life, and I am deeply grateful to him for his generosity and his friendship. I have profited from interactions 
and collaboration with numerous other colleagues, postdocs and PhD students, too many to list them all.

I thank T.~Chakraborty for encouraging me to write this review and for his patience.\\

AIPP-Data Availability Statement: Data sharing is not applicable to this article as no new data 
were created or analyzed in this study.


\begin{thebibliography}{}

\bibitem[Ruelle, 1969]{Ruelle} D.~Ruelle, \textit{Statistical Mechanics: Rigorous Results,} W.~A.~Benjamin, Inc.,
Reading MA, 1969 (second printing 1974)

\bibitem[Bratteli and Robinson, 1979-1981]{B-Rob} O.~Bratteli and D.~W.~Robinson, \textit{Operator 
Algebras and Quantum Statistical Mechanics,} vol.~1, 1979, \& vol.~2, 1981, Texts and 
Monographs in Physics, R.~Balian et al. (eds.), Springer-Verlag, New York, Berlin, Heidelberg;
(second edition 1997)

\bibitem[Simon, 1993]{BS} B.~Simon, \textit{The Statistical Mechanics of Lattice Gases,} vol.~1, Princeton 
University Press, Princeton NJ, 1993; vol.~2 in preparartion.

\bibitem[Domb et al., 1972-2001]{DGJ} C.~Domb and M.~S.~Green, \textit{Phase Transition and Critical Phenomena,}
volumes 1-6, Harcourt, Brace \& World, San Diego, New York, 1972 - 1976;\\
C. Domb and J.L.~Lebowitz, \textit{Phase Transition and Critical Phenomena,} volumes 7-20, 
Harcourt, Brace \& World, San Diego, New York, 1983-2001

\bibitem[Itzykson and Drouffe, 1989]{D-Itz} C.~Itzykson and J.-M.~Drouffe, \textit{Statistical field theory,} 
vol.~I \& II, Cambridge Monographs on Mathematical Physics, Combridge University Press, Cambridge, New York, 1989

\bibitem[Weinberg, 1995-2000]{Weinberg} S.~Weinberg, \textit{The Quantum Theory of Fields,} volumes I-III,
Cambridge University Press, Cambridge and New York, 1995 - 2000

\bibitem[Dunlop and Newman, 1974]{DN} F.~Dunlop and C.~Newman, \textit{Multicomponent Field 
Theories and Classical Rotators, } Commun.~Math.~Phys.\textbf{44}, 223-235 (1975)

\bibitem[Weiss, 1907]{Weiss} P.~Weiss, \textit{661
L’hypoth\`ese du champ mol\'eculaire et la propri\'et\'e ferromagn\'etique,} Journ. de phys. (4) \textbf{6}, 661-690 (1907);\\
see also: H.~Renker, \textit{Magnetische Untersuchungen an Legierungen der Eisengruppe oberhalb 
des Curie-Punktes,} Promotionsarbeit ETH Zurich 1913 (Referent: Herr Prof. Dr. P. Weiss, Korreferent: 
Herr Prof. Dr. A. Einstein) -- https://doi.org/10.3929/ethz-a-000092025 -- and references given there.

\bibitem[Heisenberg, 1928]{Heisenberg} W. Heisenberg, \textit{Zur Theorie des Ferromagnetismus,} 
Zeitschr.~Phys. \textbf{49}, 619-636 (1928)

\bibitem[Nambu and Jona-Lasinio, 1961]{NJL} Y. Nambu and G. Jona-Lasinio, \textit{Dynamical Model of Elementary Particles 
Based on an Analogy with Superconductivity. I,} Phys.~Rev.~\textbf{122}, 345-358 (1961)\\
Y. Nambu and G. Jona-Lasinio, \textit{Dynamical Model of Elementary Particles Based on an Analogy
with Superconductivity. II,} Phys.~Rev.~\textbf{124}, 246-254 (1961)

\bibitem[Goldstone, 1961-1962]{Goldstone} J.~Goldstone, \textit{Field Theories with ``Superconductor'' Solutions,} 
Il Nuov Cimento \textbf{XIX}, 154-164 (1961)\\
J.~Goldstone, A.~Salam and S.~Weinberg, \textit{Broken Symmetries}, Phys.~Rev.~\textbf{127}, 965-970 (1962)

\bibitem[St\"uckelberg, 1938] {Stuckelberg} 
E.~C.~G.~St\"uckelberg, \textit{Die Wechselwirkungs Kr\"afte in der Elektrodynamik und in der Feldtheorie der 
Kernkr\"afte} [\textit{The interaction forces in electrodynamics and in the field theory of nuclear forces}] (I),
Helv. Phys. Acta \textbf{11}, 225-244 (1938)\\
E.~C.~G.~St\"uckelberg, \textit{Die Wechselwirkungs Kr\"afte in der Elektrodynamik und in der Feldtheorie der 
Kernkr\"afte} (II), Helv. Phys. Acta \textbf{11}, 299-312 (1938)\\
E.~C.~G.~St\"uckelberg, \textit{Die Wechselwirkungs Kr\"afte in der Elektrodynamik und in der Feldtheorie der 
Kernkr\"afte} (III), Helv. Phys. Acta \textbf{11}, 312-328 (1938)

\bibitem[Anderson, 1962]{Anderson} P.~W.~Anderson, \textit{Plasmons, gauge invariance, and mass,} 
Phys.~Rev.~\textbf{130} (1), 439–442 (1962)

\bibitem[Higgs et al., 1964]{Higgs} F.~Englert and R.~Brout, \textit{Broken symmetry and the mass of 
gauge vector mesons,} Phys.~Rev.~Lett.~\textbf{13} (9), 321–323 (1964)\\
P.~W.~Higgs, \textit{Broken symmetries and the masses of gauge bosons,} Phys.~Rev.~Lett.~\textbf{13} (16),
508–509 (1964)\\
G.~S.~Guralnik, C.~R.~Hagen and T.~W.~B.~Kibble, \textit{Global conservation laws and massless particles,} 
Phys.~Rev.~Lett.~\textbf{13} (20), 585–587 (1964)

\bibitem['tHooft and Veltman, 1972]{'tHooft} G.~'tHooft and M.~Veltman, \textit{Regularization 
and renormalization of gauge fields,} Nucl.~Phys.~B \textbf{44} (1), 189–219 (1972)

\bibitem[Lee and Zinn-Justin, 1972]{LZ-J} B.~W.~Lee and J.~Zinn-Justin, \textit{Spontaneously Broken Gauge
Symmetries. I. Preliminaries,} Phys.~Rev.~D \textbf{5}, 3121-3137 (1972); \textit{II. Perturbation Theory and Renormalization,} Phys.~Rev.~D \textbf{5}, 3137-3155 (1972); \textit{III. Equivalence,} Phys.~Rev.~D \textbf{5},
3155-3160 (1972)

\bibitem[Fr\"ohlich et al., 1981]{FMS} J.~Fr\"ohlich, G.~Morchio and F.~Strocchi, \textit{Higgs 
phenomenon without symmetry breaking order parameter,} Nucl.~Phys.~B \textbf{190} (3), 553–582 (1981)

\bibitem[Osterwalder and Schrader, 1975]{OS} K. Osterwalder and R. Schrader, \textit{Axioms for 
Euclidean Green's functions II,} Commun.~Math.~Phys.~\textbf{42} (3), 281–305 (1975), (this paper
 is a sequel to a previous paper by the same authors from the year 1973 that contained a gap in
 the arguments)

\bibitem[Glaser, 1974]{Glaser} V. Glaser, \textit{On the equivalence of the Euclidean and 
Wightman formulation of field theory,} Commun.~Math.~Phys.~\textbf{37} (4), 257–272 (1974)

\bibitem[Simon, 1974]{Simon} B. Simon, \textit{The $P(\phi)_{2}$ Euclidean (Quantum) Field Theory,}
Princeton Series in Physics, Princeton University Press, Princeton NJ, 1974

\bibitem[Glimm and Jaffe, 1981]{GJ} J. Glimm and A. Jaffe, \textit{Quantum Physics -- 
A Functional Integral Point of View,} Spinger-Verlag, New York, Berlin, Heidelberg, 1981 ($2^{nd}$ edition
1987)

\bibitem[Fr\"ohlich, 1978]{Fr} J.~Fr\"ohlich, \textit{The Pure Phases (Harmonic Functions) of Generalized
Processes -- or: Mathematical Physics of Phase Transitions and Symmetry Breaking,} Bulletin of the AMS 
\textbf{84}, 2, 165-193 (1978)\\
J. Fr\"ohlich, \textit{Phase Transitions and Continuous Symmtery Breaking},
Lecture Notes, Erwin-Schr\"odinger Institute, Vienna 2011; (available from the author on request)

\bibitem[Fr\"ohlich et al., 1976]{FSS} J.~Fr\"ohlich, B.~Simon and T.~Spencer,  \textit{Infrared Bounds, 
Phase Transitions and Continuous Symmetry Breaking,} Commun.~Math.~Phys.~\textbf{50}, 79-95
(1976)

\bibitem[Jost, 1965]{Jost} R.~Jost, \textit{The General Theory of Quantized Fields,} Providence RI, AMS publ., 1965

\bibitem[Fr\"ohlich et al., 2020-2023]{FKSS} J.~Fr\"ohlich, A.~Knowles, B.~Schlein and V.~Sohinger, 
\textit{A Path-Integral Analysis of Interacting Bose Gases and Loop Gases,} J.~Stat.~Phys. \textbf{180}, 
810-831 (2020)\\
J.~Fr\"ohlich, A.~Knowles, B.~Schlein and V.~Sohinger, \textit{The Euclidean $\phi^{4}_2$ theory as 
a limit of an interacting Bose gas,} arXiv:2201.07632v2 [math-ph] 9 Jan 2023

\bibitem[Glimm and Jaffe, 1972]{GJ-2} J.~Glimm and A.~Jaffe, \textit{Positivity of the $\phi^{4}_3$ Hamiltonian,} Fortschritte Phys.~\textbf{21}, 327-376 (1973)

\bibitem[Feldman and Osterwalder, 1974-1977]{FO} J.~Feldman and K.~Osterwalder, \textit{The Wightman axioms and the mass gap for weakly coupled $\phi^{4}_3$ quantum field theories,} in: ``Mathematical Problems in Theoretical Physics,''
H. Araki (ed.) Berlin, Heidelberg, New York, Springer-Verlag 1975\\
 J.~Feldman and K.~Osterwalder, \textit{The Wightman axioms and the mass gap for weakly 
 coupled $\phi^{4}_3$ quantum field theories,} Ann.~Physics (NY) \textbf{97}, 80-135 (1977)
\textbf{ }

\bibitem[Park, 1977 ]{Park} Y.~M.~Park, \textit{Convergence of lattice approximation and infinite 
volumme limit in the $\big(\lambda \phi^{4} - \sigma \phi^{2} - \mu \phi\big)_{3}$ field theory,} 
J.~Math.~Phys.~\textbf{18}, 354-366 (1977)

\bibitem[Brydges et al., 1983]{BFS} D.~Brydges, J.~Fr\"ohlich and A.~Sokal, \textit{A New Proof of the 
Existence and Non-Triviality of the Continuum $\phi^{4}_{2}$ and $\phi^{4}_{3}$ Quantum Field Theories,}
Commun.~Math.~Phys.~\textbf{91}, 141-186 (1983)

\bibitem[Lee and Yang, 1952]{LY} T.~D.~Lee and C.~N.~Yang, \textit{Statistical theory of equations of state
and phase transitions II. Lattice gas and Ising model,} Phys.~Rev.~\textbf{87}, 410-419 (1952)

\bibitem[Simon and Griffiths, 1973]{SG} B.~Simon and R.~Griffiths, \textit{The  $\phi^{4}_2$-field theory as
a classical Ising model,} Commun.~Math.~Phys.~\textbf{33}, 145-164 (1973)

\bibitem[Glimm et al., 1975]{GJS} J.~Glimm, A.~Jaffe and T.~C.~Spencer, \textit{Phase transition for $\phi^{4}_2$
quantum fields,} Commun.~Math.~Phys.~\textbf{45}, 203-216 (1975)

\bibitem[Fr\"ohlich and Rodriguez, 2012]{FR} J.~Fr\"ohlich and P.-F.~Rodriguez, \textit{Some applications of the Lee-Yang theorem,} J.~Math.~Phys.~\textbf{53}, 095218-1 - 095218-15 (2012)\\
J.~Fr\"ohlich and P.-F.~Rodriguez, \textit{On Cluster Properties of Classical Ferromagnets 
in an External Magnetic Field,} J.~Stat.~Phys.~\textbf{166}, 828-840 (2017)

\bibitem[Spencer and Fr\"ohlich, 2013]{Spencer} T.~Spencer and J.~Fr\"ohlich, \textit{discussions and 
unpublished notes on the classical XY model,} IAS, Princeton, 2013-2014.

\bibitem[Symanzik, 1967]{Symanzik} K. Symanzik, \textit{Euclidean proof of the Goldstone theorem,} 
Commun.~Math.~Phys.~\textbf{6}, 228-232 (1967)

\bibitem[Buchholz et al., 1986]{BDL} D.~Buchholz, S.~Doplicher and R.~Longo, \textit{On Noether’s 
Theorem in Quantum Field Theory,} Ann.~Phys.~(NY) \textbf{170}, 1-17 (1986)

\bibitem[Hohenberg, 1967]{Hoh} P.~Hohenberg, \textit{Existence of Long-Range Order in 
One and Two Dimensions,} Phys.~Rev.~\textbf{158}, 383–386 (1967)

\bibitem[Mermin and Wagner, 1966]{MW} N.~D.~Mermin and H.~Wagner, 
\textit{Absence of Ferromagnetism or Antiferromagnetism in One- or Two-Dimensional 
Isotropic Heisenberg Models,} Phys.~Rev.~Lett.~\textbf{17}, 1133–1136 (1966);\\
N.~D.~Mermin, \textit{Absence of ordering in certain classical systems,} J.~Math.~Phys.~\textbf{6}, 1061–1064 (1967)

\bibitem[Wreszinski, 1976]{Wres} W.~Wreszinski, \textit{Goldstone’s theorem for quantum 
spin systems of finite range,} J.~Math.~Phys.~\textbf{17}, 109–111 (1976)

\bibitem[Coleman, 1973]{Coleman} S.~Coleman, \textit{There are no Goldstone Bosons
in Two Dimensions,} Commun.~Math. Phys.~\textbf{31}, 259-264 (1973)

\bibitem[Leutwyler, 1994-1997]{Leutwyler} H.~Leutwyler, \textit{Non-relativistic effective Lagrangians,} 
Phys.~Rev.~D \textbf{49}, 6, 3033-3043 (1994)\\
H.~Leutwyler, \textit{Phonons as Goldstone bosons,} Helv.~Phys.~Acta~\textbf{70}, 275-286 (1997)

\bibitem[Kosterlitz and Thouless, 1973]{BKT} M.~Kosterlitz and D.~J.~Thouless, \textit{Ordering, 
metastability and phase transitions in two-dimensional systems,} J. Phys. C \textbf{6}, 1181-1203 (1973)

\bibitem[Fr\"ohlich and Spencer, 1981]{FrSp} J.~Fr\"ohlich and T.~Spencer, \textit{The Kosterlitz-Thouless 
transition in two-dimensional abelian spin systems and the Coulomb gas,} 
Commun.~Math.~Phys.~\textbf{81}, 527-602 (1981)

\bibitem[Falco, 2012]{Falco} P.-L.~Falco, \textit{Kosterlitz-Thouless Transition Line for the 
Two Dimensional Coulomb Gas,} Commun.~Math.~Phys.~\textbf{312}, 559-609 (2012)

\bibitem[Polyakov, 1975]{Polyakov} A.~M.~Polyakov, \textit{Interaction of Goldstone Particles in Two Dimensions. Applications to Ferromagnets and Massive Yang-Mills Fields,} Phys.~Lett. \textbf{59}B, 79-81 (1975)

\bibitem[Aizenman et al., 2015]{A-DC-S} M.~Aizenman, H.~Duminil-Copin and V.~Sidoravicius, 
\textit{Random currents and continuity of Ising model's spontaneous magnetization,} Commun.~Math.~Phys.~ 
\textbf{334}, 2, 719–742 (2015)

\bibitem[Lebowitz, 1974]{Lebowitz} J.~L.~Lebowitz, \textit{GHS and other inequalities,} 
Commun.~Math.~Phys.~\textbf{35}, 87-92 (1974)

\bibitem[Glimm and Jaffe, 1974]{GJ_crit} J.~Glimm and A.~Jaffe, \textit{$\varphi^{4}_2$ quantum field 
model in the single-phase region: Differentiability of the mass and bounnds on critical exponents,} 
Phys.~Rev.~D \textbf{10}, 536-539 (1974)

\bibitem[McBryan and Rosen, 1976]{McBryan} O.~McBryan and J.~Rosen, 
\textit{Existence of the Critical Point in $\phi^{4}$ Field Theory,} Commun.~Math.~Phys.~\textbf{51}, 91-105 (1976)

\bibitem[Aizenman, 1982]{Aizenman} M.~Aizenman, \textit{Geometric analysis of $\varphi^{4}$ 
fields and Ising models. I, II,} Commun.~Math.~Phys.~\textbf{86}, 1, 1–48 (1982)

\bibitem[Fr\"ohlich, 1982]{Fr_82} J.~Fr\"ohlich, \textit{On the triviality of $\lambda \varphi^{4}_d$ theories 
and the approach to the critical point in $d > 4$ dimensions,} Nucl.~Phys.~B \textbf{200}, 2, 281–296 (1982)

\bibitem[Aizenman and Duminil-Copin, 2021]{A-DC} \textit{Marginal triviality of the scaling limits of
critical 4D Ising and $\phi^{4}_4$ models,} Ann.~Math.~\textbf{194}, 1, 163–235 (2021)

\bibitem[Ginibre, 1969]{Ginibre} J.~Ginibre, \textit{Existence of Phase Transitions for Quantum Lattice Systems,}
Commun.~Math.~Phys.~\textbf{14}, 205-234 (1969)

\bibitem[Kennedy, 1985]{Kennedy} T.~Kennedy, \textit{Long Range Order in the Anisotropic Quantum 
Ferromagnetic Heisenberg Model,} Commun.~Math.~Phys.~\textbf{100}, 447-462 (1985)

\bibitem[Fr\"ohlich et al., 1978]{FILS} J.~Fr\"ohlich, R.~Israel, E.~H.~Lieb and B.~Simon, \textit{Phase Transitions and
Reflection Positivity. I. General Theory and Long Range Lattice Models,} Commun.~Math.~Phys. 
\textbf{62}, 1-34 (1978)

\bibitem[Dyson et al., 1978]{DLS} F.~Dyson, E.~H.~Lieb and B.~Simon, \textit{Phase transitions in 
quantum spin systems with isotropic and nonisotropic interactions,} J.~Stat.~Phys.~\textbf{18}, 335-383 (1978)

\bibitem[Messager et al., 1978]{Messager} A.~Messager, S.~Miracle-Sole and C.~Pfister, \textit{Correlation 
inequalities and uniqueness of the equilibrium state for the plane rotator ferromagnetic model,} 
Commun. Math. Phys.~\textbf{58}, 1, 19–29 (1978)

\bibitem[Balaban, 1998]{Balaban} T. Balaban, \textit{The large field renormalization operation for classical 
N-vector models,} Commun.~Math.~Phys.\textbf{198}, 3, 493–534 (1998)\\
T.~Balaban and M.~O’Carroll, \textit{Low temperature properties for correlation functions 
in classical N-vector spin models,} Commun.~Math.~Phys.~\textbf{199}, 3, 493–520 (1999)

\bibitem[Bauerschmidt et al., 2021]{Bauerschmidt} R.~Bauerschmidt, N.~Crawford and T.~Helmuth, \textit{Percolation 
Transition for Random Forests in $d\geq 3$,} arXiv:2107.01878v2 [math.PR] 21 Dec 2021

\bibitem[Garban and Spencer, 2022]{GSp} Chr.~Garban and T.~Spencer, \textit{Continuous Symmetry Breaking
along the Nishimori Line,}  J.~Math.~Phys.~\textbf{63}, 9, 093302 (2022)

\bibitem[Albert et al., 2006]{Albert} C.~Albert, L.~Ferrari, J.~Fr\"ohlich and B.~Schlein, 
\textit{Magnetism and the Weiss Exhchange Field -- A Theoretical Analysis Motivated by
Recent Experiments,} J.~Stat.~Phys.~\textbf{125}, 1, 77-124 (2006)

\bibitem[Bj\"ornberg et al., 2020]{BFUe} J.~E.~Bj\"ornberg, J.~Fr\"ohlich and D.~Ueltschi, \textit{Quantum spins 
and random loops on the complete graph,} Commun.~Math.~Phys.~\textbf{375}, 1629-1663 (2020)

\bibitem[McBryan and Spencer, 1977]{McB-Sp} O.~A.~McBryan and T.~C.~Spencer, \textit{On the decay of 
correlations in SO(n)-symmetric ferromagnets,} Commun.~Math.~Phys.~\textbf{53}, 299–302 (1977)

\bibitem[Lammers, 2022]{Lammers} P.~Lammers, \textit{Height function delocalisation on cubic planar graphs,} 
Probability Theory and Related Fields \textbf{182}, 1, 531–550 (2022)\\
P.~Lammers, \textit{Bijecting the BKT transition,} arXiv:2301.06905v1 [math.PR] 17 Jan 2023

\bibitem[Fr\"ohlich and Spencer, 1983]{F-Sp} J.~Fr\"ohlich and T.~Spencer, \textit{The Berezinskii-Kosterlitz-Thouless Transition,} in: ``Scaling and Self- Similarity in Physics,'' J. Fr\"ohlich (ed.), Progress in Physics, 
Birkh\"auser, Basel, Boston, 1983

\bibitem[Guth, 1980]{Guth} A.~Guth, \textit{Existence proof of a nonconfining phase in four-dimensional 
$U(1)$ lattice gauge theory,} Phys.~Rev.~D\textbf{21}, 2291-2307 (1980)

\bibitem[Fr\"ohlich and Spencer, 1982]{F&S} J.~Fr\"ohlich and T.~Spencer, \textit{Massless Phases 
and Symmetry Restoration in Abelian Gauge Theories and Spin System,} Commun.~Math.~Phys.~\textbf{83},
411-454 (1982)

\end{thebibliography}
\end{document}